\documentclass[a4paper,11pt]{article}
\usepackage{amsmath}

\newcommand{\resection}[1]{\setcounter{equation}{0}\section{#1}}
\newcommand{\EQ}{\begin{equation}}
\newcommand{\EN}{\end{equation}}
\newcommand{\bea}{\begin{eqnarray}}
\newcommand{\eea}{\end{eqnarray}}

\begin{document}

\setcounter{page}{0} \topmargin0pt \oddsidemargin5mm \renewcommand{%
\thefootnote}{\arabic{footnote}} \newpage \setcounter{page}{0} 
\begin{titlepage}
\begin{flushright}
%OUTP-96-19S\\
SISSA 09/2005/FM
\end{flushright}
\vspace{0.5cm}
\begin{center}
{\large {\bf Form factors of descendant operators in the massive Lee-Yang
model}}\\ 
%{\large {\bf The Ising model in a magnetic field }}\\
\vspace{1.8cm}
{\large Gesualdo Delfino and Giuliano Niccoli} \\
\vspace{0.5cm}
{\em International School for Advanced Studies (SISSA)}\\
{\em via Beirut 2-4, 34014 Trieste, Italy}\\
{\em INFN sezione di Trieste}\\
{\em E-mail: delfino@sissa.it, niccoli@sissa.it}\\
%\vspace{0.5cm}
%{\large and}\\
%\vspace{0.5cm}
%{\large P. Simonetti}\\
%\vspace{0.5cm}
%{\em Department of Physics, University of Wales Swansea,\\
%Singleton Park, Swansea SA2 8PP, United Kingdom}\\
%{\em email: p.simonetti@swansea.ac.uk}\\
\end{center}
\vspace{1.2cm}

\renewcommand{\thefootnote}{\arabic{footnote}}
\setcounter{footnote}{0}

\begin{abstract}
\noindent
The form factors of the descendant operators in the massive Lee-Yang model are determined up to level 7. This is first done by exploiting the conserved quantities of the integrable theory to generate the solutions for the descendants starting from the lowest non-trivial solutions in each operator family. We then show that the operator space generated in this way, which is isomorphic to the conformal one, coincides, level by level,  with that implied by the $S$-matrix through the form factor bootstrap. The solutions we determine satisfy asymptotic conditions carrying the information about the level that we conjecture to hold for all the operators of the model.

\end{abstract}
\end{titlepage}

\newpage

\resection{Introduction}

A massive quantum field theory can be seen as the perturbation of a fixed
point (conformal) theory by relevant or marginally relevant operators. The
perturbation spoils some simple features of the conformal point, as the
power law behavior of two-point functions, but preserves some structural
properties. In particular, the operator space of the massive theory is
expected to be isomorphic to that of the conformal theory. In principle,
this expectation can be checked in $1+1$ dimensions even for non-trivial
fixed points, due to the existence of integrable quantum field theories.
What makes the check far from obvious is that the operator content at and
away from criticality is determined in two completely different ways.

On one hand, the operator content at the conformal point is dictated by the
representation theory of the Virasoro algebra underlying the infinite
dimensional conformal symmetry in two dimensions \cite{BPZ}. This reveals a
structure of operator families, each consisting of a primary operator and an
infinite tower of descendants organized into multiplets (levels) labeled by
integer spaced scaling dimensions. On the massive side, instead, integrable
quantum field theories are solved through the exact determination of the $S$%
-matrix \cite{Taniguchi}. The operators are then constructed determining
their matrix elements (form factors) on the particle states as solutions of
a set of functional equations \cite{KW,Smirnov}.

It was first shown in \cite{CM} for the thermal Ising model, and later for
more complicated models \cite{Koubek,Smirnovcounting,JMT}, that the global
counting of independent solutions of the form factor equations matches that
expected from conformal field theory. This counting procedure for the form
factor solutions, however, disentangles neither the operator families
sharing the same symmetry properties nor the levels of the descendants.

In this paper we tackle the problem of the detailed isomorphism between the
critical and off-critical operator spaces for the simplest interacting
massive theory, the Lee-Yang model. The previously available results for the
operators of this model concern the non-trivial primary operator \cite
{Alyosha} and the first non-trivial scalar descendant $T\bar{T}$ of the
identity family \cite{ttbar}. We exploit the first few quantum integrals of
motion of the theory to determine, for the two operator families of the
model and for each level up to 7, a number of independent form factor
solutions coinciding with the number predicted by conformal field theory. We
then show that the operators constructed in this way do form a basis for the
space of solutions of the form factor equations up to level seven. This is
achieved by supplementing the form factor equations with asymptotic
conditions carrying the information about the level. We also show that all
the new solutions determined in this paper automatically satisfy the
asymptotic factorization property for the descendant operators conjectured
in \cite{ttbar}.

The paper is organized as follows. In the next section we recall the
structure of the operator space at criticality as well as the $S$-matrix and
form factor language used for the integrable massive deformations. In
section~3 we specialize the discussion to the Lee-Yang model before turning
to the construction of the off-critical descendant operators through the use
of the conserved quantities in section~4. In section~5 we show the
isomorphism with the operator space spanned by the solutions of the form
factor equations, while section~6 contains few final remarks. Three
appendices conclude the paper.

\resection{Critical and off-critical operators}

The operators at a critical point undergo the general conformal field theory
classification \cite{BPZ}. A scaling operator $\Phi (x)$ is first of all
labeled by a pair $(\Delta _{\Phi },\bar{\Delta}_{\Phi })$ of conformal
dimensions which determine the scaling dimension $X_{\Phi }$ and the
euclidean spin $s_{\Phi }$ as 
\begin{eqnarray}
&&X_{\Phi }=\Delta _{\Phi }+\bar{\Delta}_{\Phi } \\
&&s_{\Phi }=\Delta _{\Phi }-\bar{\Delta}_{\Phi }.
\end{eqnarray}
Locality requires that $s_{\Phi }$ is an integer or half-integer number.
There exist operator families associated to the highest weight
representations of the Virasoro algebra 
\begin{equation}
\lbrack L_{n},L_{m}]=(n-m)L_{n+m}+\frac{c}{12}n(n^{2}-1)\delta _{n+m,0}\,.
\label{virasoro}
\end{equation}
The $L_{n}$'s generate the conformal transformations associated to the
complex variable $z=x_{1}+ix_{2}$, with the central charge $c$ labeling the
conformal theory. The same algebra, with the same value of $c$, holds for
the generators $\bar{L}_{n}$ of the conformal transformations in the
variable $\bar{z}=x_{1}-ix_{2}$. The $L_{n}$'s commute with the $\bar{L}_{m}$%
's. Each operator family consists of a primary operator $\Phi _{0}$ (which
is annihilated by all the generators $L_{n}$ and $\bar{L}_{n}$ with $n>0$)
and infinitely many descendant operators obtained through the repeated
action on the primary of the Virasoro generators. A basis in the space of
descendants is given by the operators 
\begin{equation}
L_{-i_{1}}\ldots L_{-i_{I}}\bar{L}_{-j_{1}}\ldots \bar{L}_{-j_{J}}\,\Phi _{0}
\label{descendants}
\end{equation}
with 
\begin{eqnarray}
&&0<i_{1}\leq i_{2}\leq \ldots \leq i_{I} \\
&&0<j_{1}\leq j_{2}\leq \ldots \leq j_{J}\,.
\end{eqnarray}
The levels 
\begin{equation}
(l,\bar{l})=\left( \sum_{n=1}^{I}i_{n}\,,\sum_{n=1}^{J}j_{n}\right)
\end{equation}
determine the conformal dimensions of the descendants (\ref{descendants}) in
the form 
\begin{equation}
(\Delta ,\bar{\Delta})=(\Delta _{\Phi _{0}}+l,\bar{\Delta}_{\Phi _{0}}+\bar{l%
})\,.
\end{equation}
In general the number of independent operators at level $(l,\bar{l})$ is $%
p(l)p(\bar{l})$, $p(l)$ being the number of partitions of $l$ into positive
integers. This number is reduced in presence of degenerate representations
associated to primary operators $\phi _{r,s}$ which possess a vanishing
linear combination of descendant operators (null vector) when $l$ or $\bar{l}
$ equals $rs$.

Any conformal theory possesses a primary operator corresponding to the
identity $I$ with conformal dimensions $\Delta _{I}=\bar{\Delta}_{I}=0$. All
the chiral descendants $T_{s}$ and $\bar{T}_{s}$ of the identity at level $%
(s,0)$ and $(0,s)$, respectively, are local operators satisfying the
conservation equations ($\partial \equiv \partial _{z}$, $\bar{\partial}%
\equiv \partial _{\bar{z}}$) 
\begin{equation}
\bar{\partial}T_{s}=0  \label{cftconservation1}
\end{equation}
\begin{equation}
\partial \bar{T}_{s}=0  \label{cftconservation2}
\end{equation}
associated to the infinite dimensional conformal symmetry. Since $%
L_{-1}=\partial $ and $\bar{L}_{-1}=\bar{\partial}$, the first non-vanishing
chiral descendants in the identity family are $T=L_{-2}I$ and $\bar{T}=\bar{L%
}_{-2}I$. They coincide with the non-vanishing components of the
energy-momentum tensor at criticality.

\vspace{.3cm} We can see a massive theory as the perturbation of a conformal
theory. Considering for the sake of simplicity a single relevant perturbing
operator $\varphi(x)$, we have the action 
\begin{equation}
\mathcal{A}=\mathcal{A}_{CFT}+g\int d^2x\,\varphi(x)\,.  \label{action}
\end{equation}

From the point of view of perturbation theory in $g$, a well defined
renormalization procedure can be implemented to continue the conformal
operators away from criticality \cite{Alyosha,GM}, with the result that the
operators of the perturbed theory (\ref{action}) are in one to one
correspondence with those of the conformal field theory corresponding to $%
g=0 $ \cite{Taniguchi}. A source of ambiguity in the definition of the
off-critical operators appears in presence of "resonances" \cite{Taniguchi} 
(see also \cite{FFLZZ}). An operator $\Psi $ is said to have a resonance with 
the operator $\Phi$ if the two
operators have the same spin, $s_{\Psi}=s_\Phi$, and their scaling
dimensions satisfy the condition $X_\Psi=X_\Phi+n(2-X_\varphi)$ for some
positive integer $n$. In such a case the ambiguity $\Psi\rightarrow%
\Psi+constant\,\,g^n\Phi$ mixes two operators which at criticality are
distinguished by a different scaling dimension.

The theory (\ref{action}) is integrable if operators $\Theta_s(x)$ with spin 
$s$ exist such that the conservation equations (\ref{cftconservation1}) can
be deformed into the off-critical form 
\begin{equation}
\bar{\partial}T_s=\partial\Theta_{s-2}  \label{conservation}
\end{equation}
for an infinite set of integers $s$ which is specific of the model
(similarly, (\ref{cftconservation2}) acquires a total derivative with
respect to $\bar{z}$ on the r.h.s.) \cite{Taniguchi}. Then the quantities 
\begin{equation}
Q_s=\int_{-\infty}^{+\infty}dx_1\left[T_{s+1}+\Theta_{s-1}\right]  \label{qs}
\end{equation}
with spin $s>0$, together with their counterparts $\bar{Q}_s$ with spin $-s$%
, are conserved in the massive quantum field theory. In particular, $%
\Theta_0\equiv\Theta\sim g\varphi$ is the trace of the energy-momentum
tensor, and $Q_1=P^0+P^1$ and $\bar{Q}_1=P^0-P^1$ are the light-cone
components of energy-momentum.

The existence of an infinite number of conserved quantities induces the
complete elasticity and factorization of the scattering processes, and
allows the exact determination of the $S$-matrix \cite{ZZ}. In such a
framework, the operators are constructed determining their matrix elements
on the asymptotic particle states. In order to avoid inessential
complications of the notation we consider here the case in which the
spectrum of the theory (\ref{action}) possesses a single particle of mass $%
m\sim g^{1/(2-X_{\varphi })}$. The matrix elements (form factors) 
\begin{equation}
F_{n}^{\Phi }(\theta _{1},\ldots ,\theta _{n})=\langle 0|\Phi (0)|\theta
_{1}\ldots \theta _{n}\rangle  \label{form factors}
\end{equation}
of the local operator $\Phi (x)$ between the vacuum and the $n$-particle
states\footnote{%
The rapidity variables $\theta _{k}$ parameterize the on-shell momenta of
the particles as $(p_{k}^{0},p_{k}^{1})=(m\cosh \theta _{k},m\sinh \theta
_{k})$. The generic matrix elements with particles also on the left can be
obtained by analytic continuation of (\ref{form factors}).} in the
integrable theory satisfy the functional equations \cite{KW,Smirnov} 
\begin{eqnarray}
&&F_{n}^{\Phi }(\theta _{1}+\alpha ,\ldots ,\theta _{n}+\alpha )=e^{s_{\Phi
}\alpha }F_{n}^{\Phi }(\theta _{1},\ldots ,\theta _{n})  \label{fn0} \\
&&F_{n}^{\Phi }(\theta _{1},\ldots ,\theta _{i},\theta _{i+1},\ldots ,\theta
_{n})=S(\theta _{i}-\theta _{i+1})\,F_{n}^{\Phi }(\theta _{1},\ldots ,\theta
_{i+1},\theta _{i},\ldots ,\theta _{n})  \label{fn1} \\
&&F_{n}^{\Phi }(\theta _{1}+2i\pi ,\theta _{2},\ldots ,\theta
_{n})=F_{n}^{\Phi }(\theta _{2},\ldots ,\theta _{n},\theta _{1})  \label{fn3}
\\
&&\mbox{Res}_{\theta ^{\prime }=\theta }\,F_{n+2}^{\Phi }(\theta ^{\prime }+%
\frac{i\pi }{3},\theta -\frac{i\pi }{3},\theta _{1},\ldots ,\theta
_{n})=i\Gamma \,F_{n+1}^{\Phi }(\theta ,\theta _{1},\ldots ,\theta _{n})
\label{fn2} \\
&&\mbox{Res}_{\theta ^{\prime }=\theta +i\pi }\,F_{n+2}^{\Phi }(\theta
^{\prime },\theta ,\theta _{1},\ldots ,\theta _{n})=i\left[
1-\prod_{j=1}^{n}S(\theta -\theta _{j})\right] F_{n}^{\Phi }(\theta
_{1},\ldots ,\theta _{n})\,,  \label{fn4}
\end{eqnarray}
where $S(\theta )$ is the two-particle scattering amplitude, and the
three-particle coupling $\Gamma $ is obtained from 
\begin{equation}
\mbox{Res}_{\theta =2i\pi /3}S(\theta )=i\Gamma ^{2}\,.
\end{equation}
Of course the resonance angle in the last equation as well as in (\ref{fn2})
will take more general values when spectra with particles of different
masses are considered.

The equations (\ref{fn0})--(\ref{fn4}) do not distinguish among operators
with the same spin $s_{\Phi }$. It was shown in \cite{immf} for
reflection-positive theories that the conformal dimensions determine a quite
restrictive upper bound on the asymptotic behavior of form factors at high
energies. In general, it is natural to expect a relation between the
asymptotic behavior and the level an operator of the massive theory belongs
to in the conformal limit. Indeed, the equations (\ref{fn0})--(\ref{fn4})
admit ``minimal solutions'' characterized by the mildest asymptotic
behavior, which are associated to the primary operators. The asymptotic
behavior of the non-minimal solutions, which should correspond to the
descendants, differs by integer powers of the momenta from that of the
minimal ones (see \cite{ttbar}), a pattern that obviously recalls the
integer spacing of levels in the conformal limit. Needless to say, the high
energy limit is in fact a massless limit toward the ultraviolet conformal
point.

Given a solution of the form factor equations corresponding to an operator $%
\Phi $, the conserved quantities $Q_{s}$ (as well as the $\bar{Q}_{s}$) can
be used to generate infinitely many new independent solutions. Indeed, the
commutator $[Q_{s},\Phi ]$ is the operator with conformal dimensions $%
(\Delta _{\Phi }+s,\bar{\Delta}_{\Phi })$ corresponding to the variation of $%
\Phi $ under the transformation generated by $Q_{s}$. On the other hand, the
conserved quantities annihilate the vacuum and act diagonally on the
asymptotic particle states as\footnote{%
Equations (\ref{qasymp}) and (\ref{qbarasymp}) also fix our normalization
for the conserved quantities.} 
\begin{equation}
Q_{s}|\theta _{1},\ldots ,\theta _{n}\rangle =\Lambda _{n}^{\left( s\right)
}\left( \theta _{1},..,\theta _{n}\right) |\theta _{1},\ldots ,\theta
_{n}\rangle  \label{qasymp}
\end{equation}
\begin{equation}
\bar{Q}_{s}|\theta _{1},\ldots ,\theta _{n}\rangle =\Lambda _{n}^{\left(
-s\right) }\left( \theta _{1},..,\theta _{n}\right) |\theta _{1},\ldots
,\theta _{n}\rangle \,,  \label{qbarasymp}
\end{equation}
where 
\begin{equation}
\Lambda _{n}^{\left( s\right) }\left( \theta _{1},..,\theta _{n}\right)
\equiv m^{\left| s\right| }\sum_{i=1}^{n}e^{s\theta _{i}}\,.
\label{landa-functions}
\end{equation}
Then, writing for simplicity from now on $Q_{s}\Phi $ instead of $%
[Q_{s},\Phi ]$ and $\bar{Q}_{s}\Phi $ instead of $[\bar{Q}_{s},\Phi ]$, we
have 
\begin{eqnarray}
&&F_{n}^{Q_{s}\Phi }(\theta _{1},\ldots ,\theta _{n})=-\Lambda _{n}^{\left(
s\right) }\left( \theta _{1},..,\theta _{n}\right) F_{n}^{\Phi }(\theta
_{1},\ldots ,\theta _{n})  \label{ccc-descendant-form-factors-a} \\
&&F_{n}^{\bar{Q}_{s}\Phi }(\theta _{1},\ldots ,\theta _{n})=-\Lambda
_{n}^{\left( -s\right) }\left( \theta _{1},..,\theta _{n}\right) F_{n}^{\Phi
}(\theta _{1},\ldots ,\theta _{n})\,.  \label{ccc-descendant-form-factors-b}
\end{eqnarray}
It is straightforward to check that these expressions satisfy the general
form factor equations (\ref{fn0})--(\ref{fn3}) for operators with spin $%
s_{\Phi }+s$ and $s_{\Phi }-s$, respectively. Equation (\ref{fn4}) leads to
the condition 
\begin{equation}
e^{s\theta }+e^{s(\theta +i\pi )}=0\,,
\end{equation}
showing that the values of the spin $s$ of the conserved quantities need to
be odd. Finally, if $\Gamma \neq 0$, equation (\ref{fn2}) gives the
condition 
\begin{equation}
e^{is\pi /3}+e^{-is\pi /3}=1\,,
\end{equation}
showing that, in a theory with a single particle $A$ in the spectrum and
admitting the fusion $AA\rightarrow A$, the allowed values for the spin of
the conserved quantities are 
\begin{equation}
s=6n\pm 1\,,\hspace{1cm}n=0,\pm 1,\pm 2\ldots \,\,.  \label{spins}
\end{equation}

\resection{\label{gamma-2}The Lee-Yang model}

\subsection{Critical point}

The Lee-Yang model is the quantum field theory associated to the edge
singularity of the zeros of the partition function of the Ising model in an
imaginary magnetic field \cite{YL,LY,Fisher}. The critical point is
described by the simplest conformal field theory \cite{Cardy}, i.e. the
minimal model $\mathcal{M}_{2,5}$, with central charge $c=-22/5$, possessing
only two local primary operators: the identity $I=\phi_{1,1}=\phi_{1,4}$
with conformal dimensions $(0,0)$, and the operator $\varphi=\phi_{1,2}=%
\phi_{1,3}$ with conformal dimensions $(-1/5,-1/5)$. The negative values of
the central charge and of $X_\varphi$ show that the model does not satisfy
reflection-positivity.

The degenerate nature of the primary fields reduces the dimensionality of
the space of the level $(l,\bar{l})$ descendants of the primary $\phi $ to $%
d_{\phi }(l)d_{\phi }(\bar{l})$, with $d_{\phi }(n)$ generated by the
rescaled character 
\begin{equation}
\chi _{\phi }(q)=\sum_{n=0}^{\infty }d_{\phi }(n)q^{n}\,.
\end{equation}
The two characters for the Lee-Yang model are \cite{Feigen-Fuchs,R-C,Christe}
\begin{equation*}
\chi _{I}(q)=\prod_{n=0}^{\infty }\frac{1}{\left( 1-q^{2+5n}\right) \left(
1-q^{3+5n}\right) }
\end{equation*}
\begin{equation*}
\chi _{\varphi }(q)=\prod_{n=0}^{\infty }\frac{1}{\left( 1-q^{1+5n}\right)
\left( 1-q^{4+5n}\right) }\,.
\end{equation*}
The dimensionalities of the first few levels can then be read from the
expansions 
\begin{equation*}
\chi _{I}(q)=1\hspace{0.02in}+q^{2}+q^{3}\hspace{0.02in}+\hspace{0.02in}q^{4}%
\hspace{0.02in}+\hspace{0.02in}q^{5}\hspace{0.02in}+\hspace{0.02in}2q^{6}+%
\hspace{0.02in}2q^{7}+\hspace{0.02in}3q^{8}+O(q^{9})
\end{equation*}
\begin{equation*}
\chi _{\varphi
}(q)=1+q+q^{2}+q^{3}+2q^{4}+2q^{5}+3q^{6}+3q^{7}+4q^{8}+O(q^{9})\,.
\end{equation*}
A basis for the chiral descendants up to level 7 is given in table~1.

\begin{table}[tbp]
\begin{center}
\begin{tabular}{|l||l|l|}
\hline
$\text{Level of the descendant}$ & $\hspace{0.1in}I$ & $\hspace{0.1in}%
\varphi $ \\ \hline\hline
\ \ \ \ \ \ \ \ \ \ \ \ \ \ \ $\left( 1,0\right) $ & $\hspace{0.1in}0$ & $%
\hspace{0.1in}\partial \varphi $ \\ \hline
\ \ \ \ \ \ \ \ \ \ \ \ \ \ \ $\left( 2,0\right) $ & $\hspace{0.1in}T$ & $%
\hspace{0.1in}\partial ^{2}\varphi $ \\ \hline
\ \ \ \ \ \ \ \ \ \ \ \ \ \ \ $\left( 3,0\right) $ & $\hspace{0.1in}\partial
T$ & $\hspace{0.1in}\partial ^{3}\varphi $ \\ \hline
\ \ \ \ \ \ \ \ \ \ \ \ \ \ \ $\left( 4,0\right) $ & $\hspace{0.1in}\partial
^{2}T$ & $\hspace{0.1in}\partial ^{4}\varphi \hspace{0.05in};\hspace{0.05in}%
L_{-4}\varphi $ \\ \hline
\ \ \ \ \ \ \ \ \ \ \ \ \ \ \ $\left( 5,0\right) $ & $\hspace{0.1in}\partial
^{3}T$ & $\hspace{0.1in}\partial ^{5}\varphi \hspace{0.05in};\hspace{0.05in}%
\partial L_{-4}\varphi $ \\ \hline
\ \ \ \ \ \ \ \ \ \ \ \ \ \ \ $\left( 6,0\right) $ & $\hspace{0.1in}\partial
^{4}T\hspace{0.05in};\hspace{0.05in}L_{-4}T$ & $\hspace{0.1in}\partial
^{6}\varphi \hspace{0.05in};\hspace{0.05in}\partial ^{2}L_{-4}\varphi 
\hspace{0.05in};\hspace{0.05in}L_{-6}\varphi $ \\ \hline
\ \ \ \ \ \ \ \ \ \ \ \ \ \ \ $\left( 7,0\right) $ & $\hspace{0.1in}\partial
^{5}T\hspace{0.05in};\hspace{0.05in}\partial L_{-4}T$ & $\hspace{0.1in}%
\partial ^{7}\varphi \hspace{0.05in};\hspace{0.05in}\partial
^{3}L_{-4}\varphi \hspace{0.05in};\hspace{0.05in}\partial L_{-6}\varphi $ \\ 
\hline
\end{tabular}
\end{center}
\caption{Basis for the left chiral descendants of the two primary fields in
the Lee-Yang model up to level 7 in terms of Virasoro generators. The
corresponding basis for the right descendants is obtained changing $\partial 
$, $L_{-n}$ and $T$ into $\bar{\partial}$, $\bar{L}_{-n}$ and $\bar{T}$,
respectively.}
\end{table}

\begin{table}[tbp]
\begin{center}
\begin{tabular}{|l||l|l|}
\hline
$\text{Level of the descendant}$ & $\hspace{0.1in}I$ & $\hspace{0.1in}%
\varphi $ \\ \hline\hline
\ \ \ \ \ \ \ \ \ \ \ \ \ \ \ $\left( 1,0\right) $ & $\hspace{0.1in}0$ & $%
\hspace{0.1in}\partial \varphi $ \\ \hline
\ \ \ \ \ \ \ \ \ \ \ \ \ \ \ $\left( 2,0\right) $ & $\hspace{0.1in}T$ & $%
\hspace{0.1in}\partial ^{2}\varphi $ \\ \hline
\ \ \ \ \ \ \ \ \ \ \ \ \ \ \ $\left( 3,0\right) $ & $\hspace{0.1in}\partial
T$ & $\hspace{0.1in}\partial ^{3}\varphi $ \\ \hline
\ \ \ \ \ \ \ \ \ \ \ \ \ \ \ $\left( 4,0\right) $ & $\hspace{0.1in}\partial
^{2}T$ & $\hspace{0.1in}\partial ^{4}\varphi \hspace{0.05in};\hspace{0.05in}%
R_{4}\varphi $ \\ \hline
\ \ \ \ \ \ \ \ \ \ \ \ \ \ \ $\left( 5,0\right) $ & $\hspace{0.1in}\partial
^{3}T$ & $\hspace{0.1in}\partial ^{5}\varphi \hspace{0.05in};\hspace{0.05in}%
Q_{5}\varphi $ \\ \hline
\ \ \ \ \ \ \ \ \ \ \ \ \ \ \ $\left( 6,0\right) $ & $\hspace{0.1in}\partial
^{4}T\hspace{0.05in};\hspace{0.05in}S_{4}T$ & $\hspace{0.1in}\partial
^{6}\varphi \hspace{0.05in};\hspace{0.05in}\partial Q_{5}\varphi \hspace{%
0.05in};\hspace{0.05in}R_{6}\varphi $ \\ \hline
\ \ \ \ \ \ \ \ \ \ \ \ \ \ \ $\left( 7,0\right) $ & $\hspace{0.1in}\partial
^{5}T\hspace{0.05in};\hspace{0.05in}Q_{5}T$ & $\hspace{0.1in}\partial
^{7}\varphi \hspace{0.05in};\hspace{0.05in}\partial ^{2}Q_{5}\varphi \hspace{%
0.05in};\hspace{0.05in}Q_{7}\varphi $ \\ \hline
\end{tabular}
\end{center}
\caption{Basis for the left chiral descendants of the two primary fields in
the Lee-Yang model up to level 7 exploiting the conserved quantities. An
analogous basis exists for the right chiral descendants.}
\end{table}

At the conformal point the conserved quantities $Q_{s}$ are combinations of
the Virasoro generators $L_{n}$ with dimensions $(s,0)$, while the $\bar{Q}%
_{s}$ are combinations of the $\bar{L}_{n}$ with dimensions $(0,s)$.
However, since $Q_s$ and $\bar{Q}_s$ are conserved in the massive theory and
act diagonally on the particle states, they are much more convenient than
the Virasoro generators for labeling the descendant operators in view of the
off-critical continuation. This is why we exploit the existence of conserved
quantities with spin $5$ and $7$ in the Lee-Yang model to switch from the
basis of table~1 to that given in table~2. Here the descendant operators $%
S_{4}T$, $R_{4}\varphi $ and $R_{6}\varphi $ are defined through the
relations 
\begin{eqnarray}
&&Q_{5}T=\partial S_{4}T  \label{i6} \\
&&Q_{5}\varphi =\partial R_{4}\varphi  \label{r4} \\
&&Q_{7}\varphi =\partial R_{6}\varphi \,\,.  \label{r6}
\end{eqnarray}
The descendant operators $\bar{S}_{4}\bar{T}$, $\bar{R}_{4}\varphi $ and $%
\bar{R}_{6}\varphi $ defined through the relations 
\begin{eqnarray}
&&\bar{Q}_{5}\bar{T}=\bar{\partial}\bar{S}_{4}\bar{T}  \label{sb4} \\
&&\bar{Q}_{5}\varphi =\bar{\partial}\bar{R}_{4}\varphi  \label{rb4} \\
&&\bar{Q}_{7}\varphi =\bar{\partial}\bar{R}_{6}\varphi  \label{bar6}
\end{eqnarray}
appear in the right chiral basis.

The linear independence between $\partial ^{5}T$ and $Q_{5}T$ ensures that
we can write $Q_{5}T\sim \partial ^{5}T+a\partial L_{-4}T$, with $a\neq 0$.
Then (\ref{i6}) ensures that $S_{4}T$ contains a non-vanishing component of $%
L_{-4}T$. Similar considerations apply to the other operators we have
defined through (\ref{i6})--(\ref{bar6}). Then we write 
\begin{eqnarray*}
&S_{4}\sim \partial ^{4}+aL_{-4}\,,&\bar{S}_{4}\sim \bar{\partial}^{4}+a\bar{%
L}_{-4} \\
&R_{4}\sim \partial ^{4}+bL_{-4}\,,&\bar{R}_{4}\sim \bar{\partial}^{4}+b\bar{%
L}_{-4} \\
&R_{6}\sim \partial ^{6}+c\partial ^{2}L_{-4}+dL_{-6}\,,\hspace{1cm}&\bar{R}%
_{6}\sim \bar{\partial}^{6}+c\bar{\partial}^{2}L_{-4}+d\bar{L}_{-6}\,,
\end{eqnarray*}
with $a$, $b$ and $d$ different from zero.

\subsection{Scaling limit}

The off-critical Lee-Yang model is obtained perturbing the conformal field
theory $\mathcal{M}_{2,5}$ with the primary operator $\varphi $, which is
the only non-trivial relevant operator in the theory. The massive model is
integrable as a consequence of the general results of \cite{Taniguchi} about
perturbed conformal field theories. The mass spectrum contains only one
particle $A$ and the exact on-shell solution is provided by the $S$-matrix 
\cite{CMyanglee} 
\begin{equation}
S(\theta )=\frac{\tanh \frac{1}{2}\left( \theta +\frac{2i\pi }{3}\right) }{%
\tanh \frac{1}{2}\left( \theta -\frac{2i\pi }{3}\right) }\,,  \label{s}
\end{equation}
where $\theta $ is the rapidity difference of the two colliding particles.
The bound state pole located at $\theta =2i\pi /3$ implies the fusion
property $AA\rightarrow A$, so that the model possesses conserved quantities
with the spins given in (\ref{spins}).

The trace of the energy-momentum tensor $\Theta =T_{\mu }^{\mu }/4$ is
proportional to the perturbing operator $\varphi $ and from now on we refer
to $\Theta $ instead of $\varphi $ in the off-critical theory. The form
factors of $\Theta $ were determined in \cite{Smirnovreduction,Alyosha} (see
also \cite{Koubek}) and can be written as 
\begin{equation}
F_{n}^{\Theta }(\theta _{1},\ldots ,\theta _{n})=U_{n}^{\Theta }(\theta
_{1},\ldots ,\theta _{n})\prod_{i<j}\frac{F_{min}(\theta _{i}-\theta _{j})}{%
\cosh \frac{\theta _{i}-\theta _{j}}{2}\left[ \cosh (\theta _{i}-\theta
_{j})+\frac{1}{2}\right] }\text{.}  \label{fn}
\end{equation}
Here the factors in the denominator introduce the bound state and
annihilation poles prescribed by (\ref{fn2}) and (\ref{fn4}), while 
\begin{equation}
F_{min}(\theta )=-i\sinh \frac{\theta }{2}\,\exp \left\{ 2\int_{0}^{\infty }%
\frac{dt}{t}\frac{\cosh \frac{t}{6}}{\cosh \frac{t}{2}\sinh t}\sin ^{2}\frac{%
(i\pi -\theta )t}{2\pi }\right\}  \label{fmin}
\end{equation}
is a solution of the equations 
\begin{equation}
F(\theta )=S(\theta )F(-\theta )
\end{equation}
\begin{equation}
F(\theta +2i\pi )=F(-\theta )
\end{equation}
free of zeros and poles for Im$\theta \in (0,2\pi )$ and behaving as 
\begin{equation}
F_{min}(\theta )\sim e^{|\theta |}  \label{asyp-fmin}
\end{equation}
when $|\theta |\rightarrow \infty $. Finally, $U_{n}^{\Theta }(\theta
_{1},\ldots ,\theta _{n})$ are the following entire functions of the
rapidities, symmetric and (up to a factor $(-1)^{n-1}$) $2\pi i$-periodic in
all $\theta _{j}$'s: 
\begin{equation}
U_{n}^{\Theta }\left( \theta _{1},..,\theta _{n}\right) =H_{n}\left( \frac{1%
}{\sigma _{n}^{(n)}}\right) ^{\left( n-1\right) /2}Q_{n}^{\Theta }\left(
\theta _{1},..,\theta _{n}\right) \,.  \label{un}
\end{equation}
Here $\sigma _{i}^{(n)}$ are the elementary symmetric polynomials generated
by 
\begin{equation}
\prod_{i=1}^{n}(x+x_{i})=\sum_{k=0}^{n}x^{n-k}\sigma _{k}^{(n)}(x_{1},\ldots
,x_{n})\text{,}
\end{equation}
with $x_{i}\equiv e^{\theta _{i}}$, $H_{n}=i^{n^{2}}\left( 3/4\right)
^{n/4}\gamma ^{n\left( n-2\right) }$, 
\begin{equation}
\gamma =\exp \left\{ 2\int_{0}^{\infty }\frac{dt}{t}\,\frac{\sinh \frac{t}{2}%
\sinh \frac{t}{3}\sinh \frac{t}{6}}{\sinh ^{2}t}\right\} \,,
\end{equation}
and $Q_{n}^{\Theta }(\theta _{1},\ldots ,\theta _{n})$ is the determinant 
\begin{equation}
Q_{n}^{\Theta }(\theta _{1},\ldots ,\theta _{n})=-\frac{\pi m^{2}}{4\sqrt{3}}%
\,det\left\| M_{i,j}^{(n)}\right\|  \label{thetan}
\end{equation}
of the $(n-1)\times (n-1)$ matrix with entries 
\begin{equation}
M_{i,j}^{(n)}=\frac{\sin (2(i-j+1)\frac{\pi }{3})}{\sin \frac{2\pi }{3}}%
\,\sigma _{2i-j}^{(n)}\,.
\end{equation}

The conservation equations 
\begin{equation}
\bar{\partial}T=\partial \Theta \,,\hspace{0.3in}\partial \bar{T}=\bar{%
\partial}\Theta
\end{equation}
allow the determination of the form factors of the other components of the
energy-momentum tensor in the form 
\begin{eqnarray}
&&F_{n}^{T}(\theta _{1},\ldots ,\theta _{n})=-\frac{\sigma _{n}^{(n)}\sigma
_{1}^{(n)}}{\sigma _{n-1}^{(n)}}\,F_{n}^{\Theta }(\theta _{1},\ldots ,\theta
_{n})  \label{T-descendant} \\
&&F_{n}^{\bar{T}}(\theta _{1},\ldots ,\theta _{n})=-\,\frac{\sigma
_{n-1}^{(n)}}{\sigma _{1}^{(n)}\sigma _{n}^{(n)}}F_{n}^{\Theta }(\theta
_{1},\ldots ,\theta _{n})  \label{Tbar-descendant}
\end{eqnarray}
for $n>0$. Of course $\langle T\rangle =\langle \bar{T}\rangle =0$ as for
any operator with non-zero spin. The factors in the denominator of the last
two equations do not introduce unwanted singularities since the functions $%
Q_{n}^{\Theta }$ have the property of factorizing $\sigma _{1}^{(n)}$ and $%
\sigma _{n-1}^{(n)}$, respectively the eigenvalues\footnote{%
More precisely 
\begin{equation*}
\partial |\theta _{1},\ldots ,\theta _{n}\rangle =-im\sigma
_{1}^{(n)}|\theta _{1},\ldots ,\theta _{n}\rangle \hspace{0.1in};\hspace{%
0.1in}\bar{\partial}|\theta _{1},\ldots ,\theta _{n}\rangle =im\bar{\sigma}%
_{1}^{(n)}|\theta _{1},\ldots ,\theta _{n}\rangle
\end{equation*}
where 
\begin{equation*}
\bar{\sigma}_{i}^{(n)}\equiv \sigma _{n-i}^{(n)}/\sigma _{n}^{(n)}\text{.}
\end{equation*}
} of $\partial $ and $\bar{\partial}$ on the $n$-particle asymptotic states.

\resection{Off-critical descendants from conservation laws}

In this section we show how the form factors of all the descendant operators
in the massive Lee-Yang model up to level 7 can be obtained starting from
the known solutions for $\Theta$, $T$, $\bar{T}$ and $T\bar{T}$ exploiting
the conserved quantities of the integrable model.

All the solutions we find in this way automatically satisfy the following
asymptotic properties that we conjecture to hold for all the operators of
the theory\footnote{%
We show in appendix~B that these properties hold for all the descendant
operators obtained acting on a primary with conserved quantities.}:

i) the form factors of a level $(l,\bar{l})$ non-chiral descendant $\Phi
_{(l,\bar{l})}$ of the primary $\Phi _{0}$ are homogeneous functions, with
respect to the variables $x_{i}\equiv e^{\theta _{i}}$, of degree $l-\bar{l}$
(eq. (\ref{fn0})), furthermore behaving as 
\begin{equation}
F_{n}^{\Phi _{(l,\bar{l})}}\left( \theta _{1}+\alpha ,..,\theta _{k}+\alpha
,\theta _{k+1},...,\theta _{n}\right) \sim e^{l\alpha }
\label{level-descendant}
\end{equation}
for\ $\alpha\rightarrow +\infty$, $n>1\hspace{0.08in}$and$\hspace{0.08in}%
1\leq k\leq n-1$. This equation holds also for the chiral descendants of $%
\Theta $, while the chiral descendants of $I$ of level $\left( l,0\right) $
and $(0,\bar{l})$ have form factors with homogeneous degree $l$ and $-\bar{l}
$, respectively, and behave as 
\begin{equation}
F_{n}^{I_{\left( l,0\right) }}\left( \theta _{1}+\alpha ,..,\theta
_{k}+\alpha ,\theta _{k+1},...,\theta _{n}\right) \sim e^{\left( l-1\right)
\alpha }\,,\hspace{0.08in}F_{n}^{I_{(0,\bar{l})}}\left( \theta _{1}+\alpha
,..,\theta _{k}+\alpha ,\theta _{l+1},...,\theta _{n}\right) \sim e^{-\alpha
}  \label{chiral-descendants-Identity}
\end{equation}
for\ $\alpha\rightarrow +\infty$, $n>1\hspace{0.08in}$and$\hspace{0.08in}%
1\leq k\leq n-1$;

ii) let us denote by $\mathcal{L}_{l}\bar{\mathcal{L}}_{\bar{l}}\Phi _{0}$ a
level $(l,\bar{l})$ descendant which in the critical limit assumes the form (%
\ref{descendants}) with $\mathcal{L}_{l}$ ($\bar{\mathcal{L}}_{\bar{l}}$)
accounting for the product of $L_{-i}$ ($\bar{L}_{-j}$). Then the
factorization property 
\begin{equation}
\lim_{\alpha \rightarrow +\infty }e^{-l\alpha }F_{n}^{\mathcal{L}_{l}\bar{%
\mathcal{L}}_{\bar{l}}\Phi _{0}}\left( \theta _{1}+\alpha ,..,\theta
_{k}+\alpha ,\theta _{k+1},..,\theta _{n}\right) =\frac{1}{\left\langle \Phi
_{0}\right\rangle }F_{k}^{\mathcal{L}_{l}\Phi _{0}}\left( \theta
_{1},..,\theta _{k}\right) F_{n-k}^{\bar{\mathcal{L}}_{\bar{l}}\Phi
_{0}}\left( \theta _{k+1},..,\theta _{n}\right)  \label{Conjecture-2}
\end{equation}
holds for $k=0,1,\ldots ,n$.

\subsection{Descendants of $\Theta$}

The form factors of the descendants of $\Theta $ obtained acting on the
primary with the conserved quantities are defined without ambiguity by $%
\left( \ref{ccc-descendant-form-factors-a}\right) $ and $\left( \ref
{ccc-descendant-form-factors-b}\right) $. Then we are left with the problem
of fixing the form factors of the operators obtained acting on $\Theta $
with $R_{4}$ , $R_{6}$ and their negative spin counterparts. We know from (%
\ref{r4}), (\ref{r6}), (\ref{rb4}) and (\ref{bar6}) that at criticality
these operators differ by derivatives from operators obtained through the
action of conserved quantities. Since the form factors of $\Theta $
factorize the eigenvalues of $\partial \bar{\partial}$, this property is
automatically satisfied also in the massive theory. Indeed, stripping off in
each case the appropriate derivative eigenvalue we can write\vspace{0.07in} 
\begin{equation*}
\begin{array}{l}
\begin{tabular}{ll}
$\overset{}{F_{n}^{R_{4}\Theta }=\left( im\sigma _{1}^{(n)}\right)
^{-1}F_{n}^{Q_{5}\Theta }},$ & \ \ \ \ \ \ \ \ \ \ \ \ \ \ \ \ \ $\overset{}{%
F_{n}^{\bar{R}_{4}\Theta }=\left( -im\bar{\sigma}_{1}^{(n)}\right)
^{-1}F_{n}^{\bar{Q}_{5}\Theta }}$%
\end{tabular}
\\ 
\\ 
\begin{tabular}{ll}
$\overset{}{F_{n}^{R_{6}\Theta }=\left( im\sigma _{1}^{(n)}\right)
^{-1}F_{n}^{Q_{7}\Theta }},$ & \ \ \ \ \ \ \ \ \ \ \ \ \ \ \ \ \ $\overset{}{%
F_{n}^{\bar{R}_{6}\Theta }=\left( -im\bar{\sigma}_{1}^{(n)}\right)
^{-1}F_{n}^{\bar{Q}_{7}\Theta }}\,$%
\end{tabular}
\\ 
\\ 
\begin{tabular}{ll}
$\overset{}{F_{n}^{R_{4}\bar{R}_{4}\Theta }=\left( m^{2}\sigma _{1}^{(n)}%
\bar{\sigma}_{1}^{(n)}\right) ^{-1}F_{n}^{Q_{5}\bar{Q}_{5}\Theta }},$ & $%
\overset{}{\ \ \ \ \ \ F_{n}^{R_{4}\bar{R}_{6}\Theta }=\left( m^{2}\sigma
_{1}^{(n)}\bar{\sigma}_{1}^{(n)}\right) ^{-1}F_{n}^{Q_{5}\bar{Q}_{7}\Theta }}
$%
\end{tabular}
\\ 
\\ 
\begin{tabular}{ll}
$\overset{}{F_{n}^{R_{6}\bar{R}_{6}\Theta }=\left( m^{2}\sigma _{1}^{(n)}%
\bar{\sigma}_{1}^{(n)}\right) ^{-1}F_{n}^{Q_{7}\bar{Q}_{7}\Theta }},$ & \ \
\ \ \ \ $\overset{}{F_{n}^{R_{6}\bar{R}_{4}\Theta }=\left( m^{2}\sigma
_{1}^{(n)}\bar{\sigma}_{1}^{(n)}\right) ^{-1}F_{n}^{Q_{7}\bar{Q}_{5}\Theta }.%
}$%
\end{tabular}
\end{array}
\end{equation*}
\vspace{0.07in}

The form factors of $\Theta $ given in the previous section satisfy the
asymptotic factorization property ($k=0,1,\ldots ,n$) 
\begin{equation}
\lim_{\alpha \rightarrow +\infty }F_{n}^{\Theta }\left( \theta _{1}+\alpha
,..,\theta _{k}+\alpha ,\theta _{k+1},..,\theta _{n}\right) =\frac{1}{%
\langle \Theta \rangle }F_{k}^{\Theta }(\theta _{1},..,\theta
_{k})F_{n-k}^{\Theta }\left( \theta _{k+1},..,\theta _{n}\right)
\label{clustertheta}
\end{equation}
discussed in \cite{DSC}, with 
\begin{equation}
\langle \Theta \rangle=F_{0}^{\Theta }=-\frac{\pi m^{2}}{4\sqrt{3}}\,.
\label{vevtheta}
\end{equation}
This value of the vacuum expectation value was originally obtained in \cite
{TBA} using the thermodynamic Bethe ansatz.

The factorization properties of the eigenvalues of the conserved quantities
given in appendix B then ensure that the solutions determined above satisfy (%
\ref{Conjecture-2}) with $\Phi _{0}=\Theta $. This in turn implies the
asymptotic behavior (\ref{level-descendant}).

\subsection{Descendants of the identity}

The matrix elements of the chiral descendants of $I$ obtained through
conserved charges are again fixed by $\left( \ref
{ccc-descendant-form-factors-a}\right) $ and $\left( \ref
{ccc-descendant-form-factors-b}\right) $ together with the solution for $T$
and $\bar{T}$ given in the previous section. Concerning the remaining chiral
operators $S_{4}T$ and $\bar{S}_{4}\bar{T}$, we observe that the form
factors of $T$ and $\bar{T}$ factorize the eigenvalues of $\partial ^{2}$
and $\bar{\partial}^{2}$, respectively. We can then exploit (\ref{i6}) and (%
\ref{sb4}) to write 
\begin{equation}
F_{n}^{S_{4}T}=\left( im\sigma _{1}^{(n)}\right) ^{-1}F_{n}^{Q_{5}T}\text{ \ 
},\text{ \ }F_{n}^{\bar{S}_{4}\bar{T}}=\left( -im\bar{\sigma}%
_{1}^{(n)}\right) ^{-1}F_{n}^{\bar{Q}_{5}\bar{T}}\,.
\end{equation}

We can now turn to the problem of identifying the off-critical continuation
of the non-chiral descendants of $I$. The first of such operators is $T\bar{T%
}$. It was shown in \cite{Sasha} that this operator can be conveniently
defined away from criticality as 
\begin{equation}
T\bar{T}(x)=\lim_{x^{\prime }\rightarrow x}[T(x)\bar{T}(x^{\prime })-%
\mathcal{R}_{T\bar{T}}(x,x^{\prime })]\,\text{,}  \label{regularized}
\end{equation}
where 
\begin{equation}
\mathcal{R}_{T\bar{T}}(x,x^{\prime })=\Theta (x)\Theta (x^{\prime })+%
\mbox{derivative terms}  \label{regttbar}
\end{equation}
regularizes the divergences arising in the limit away from criticality. The
matrix elements of $T\bar{T}$ in the Lee-Yang model have been identified in 
\cite{ttbar} and read 
\begin{equation}
F_{n}^{T\bar{T}}=\frac{\langle \Theta \rangle }{m^{4}}\,F_{n}^{\partial ^{2}%
\bar{\partial}^{2}\Theta }+\langle \Theta \rangle
^{2}\,F_{n}^{K_{3}}-2\,\langle \Theta \rangle F_{n}^{\Theta }+\langle \Theta
\rangle ^{2}\,\,\delta _{n,0}+c\,F_{n}^{\partial \bar{\partial}\Theta }\,.
\label{fnttbar}
\end{equation}
The kernel solution (see next section) $F_{n}^{{K}_{3}}$ is given explicitly
in \cite{ttbar} up to $n=9$. The operator $\partial \bar{\partial}\Theta $
is resonant with $T\bar{T}$, so that the coefficient $c$ in (\ref{fnttbar})
is intrinsically ambiguous.

Going to higher non-chiral descendants of the identity, consider those
operators which are obtained as regularized products of conserved quantities
acting on $T$ and $\bar{T}$. It is shown in appendix A that the subtraction
which regularizes the operator (\ref{regularized}) regularizes also these
composite operators. We exploit, however, the possibility of adding resonant
terms $\rho _{k}$, and write 
\begin{equation}
Q_{s}\bar{Q}_{r}T\bar{T}(x)=Q_{s}\bar{Q}_{r}\left( T\bar{T}\right)
(x)+\sum_{k}c_{k}\rho _{k}(x)\,,  \label{ccc-descendant-ttbar}
\end{equation}
or in terms of form factors, 
\begin{equation}
F_{n}^{Q_{s}\bar{Q}_{r}T\bar{T}}=\Lambda _{n}^{\left( s\right) }\Lambda
_{n}^{\left( -r\right) }F_{n}^{T\bar{T}}+\sum_{k}c_{k}F_{n}^{\rho _{k}}\,.
\label{form factors-ccc-descendant-ttbar}
\end{equation}
The role of the resonances is the following. We know from table~2 and
equations (\ref{i6})-(\ref{bar6}) that at criticality some of the operators
we are considering are derivatives of operators living one level below. We
expect this property to continue to hold in the massive theory and have seen
how this is automatically achieved for the descendants of $\Theta $ and the
chiral descendants of $I$. Concerning the non-chiral descendants of $I$, the
form factor solution for $T\bar{T}$ does not factorize any derivative
eigenvalue (see \cite{ttbar}), and this property is not guaranteed a priori.
Remarkably, however, it can be recovered by fine tuning the coefficients of
some of the resonant terms in the last equation.

To see this, observe first that, since $T\bar{T}$ and $\Theta $ have
conformal dimensions $(2,2)$ and $(1,1)$, respectively, the level $(p,q)$
descendant of $I$ is resonant with the level $(p-1,q-1)$ descendants of $%
\Theta $ for $p$ and $q$ larger than 1. Letting aside for the time being the
operators containing $S_{4}$ and $\bar{S}_{4}$, for $p$ and $q$ less than 7
the above requirement only forces the coefficients of resonances like $R_{4}%
\bar{\partial}\Theta $, $\partial \bar{R}_{4}\Theta $ and $R_{4}\bar{R}%
_{4}\Theta $ (which do not contain the required derivative terms) to vanish.

A more interesting situation arises when considering the descendants of $I$
containing $Q_{5}$ or $\bar{Q}_{5}$ acting on $T\bar{T}$. At level $(7,2)$
we have two operators, $Q_{5}T\bar{T}$ and $\partial ^{5}T\bar{T}$, which
away from criticality are resonant with the space of level $(6,1)$
descendants of $\Theta $ spanned by $Q_{5}\partial \bar{\partial}\Theta $, $%
\partial ^{6}\bar{\partial}\Theta $ and $R_{6}\bar{\partial}\Theta $. The
latter operator, with form factors 
\begin{equation}
F_{n}^{R_{6}\bar{\partial}\Theta }=\frac{\bar{\sigma}_{1}^{(n)}}{\sigma
_{1}^{(n)}}\Lambda _{n}^{\left( 7\right) }F_{n}^{\Theta }\,,
\end{equation}
is the only one among the resonant operators which does not contain a
derivative with respect to $z$. The requirement that also away from
criticality the operators $Q_{5}T\bar{T}$ and $\partial ^{5}T\bar{T}$ are
derivatives with respect to $z$ of the two level $(6,2)$ descendants of $I$
is satisfied with the following unique choice of the coefficients of $R_{6}%
\bar{\partial}\Theta $ 
\begin{eqnarray}
&&\partial ^{5}T\bar{T}(x)=\partial ^{5}\left( T\bar{T}\right) (x)
\label{t5} \\
&&Q_{5}T\bar{T}(x)=Q_{5}\left( T\bar{T}\right) (x)-\frac{5}{7}R_{6}\bar{%
\partial}\Theta (x)\,.  \label{Q5-ttbar}
\end{eqnarray}
Notice that, since the composite operator $T\bar{T}$ already includes the
resonant term $\partial \bar{\partial}\Theta $ in its definition (see (\ref
{fnttbar})), the resonances $Q_{5}\partial \bar{\partial}\Theta $ and $%
\partial ^{6}\bar{\partial}\Theta $ are automatically included in the space
spanned by (\ref{t5}) and (\ref{Q5-ttbar}). The same analysis gives 
\begin{eqnarray}
&&\bar{\partial}^{5}T\bar{T}(x)=\bar{\partial}^{5}\left( T\bar{T}\right) (x)
\\
&&\bar{Q}_{5}T\bar{T}(x)=\bar{Q}_{5}\left( T\bar{T}\right) (x)-\frac{5}{7}%
\partial \bar{R}_{6}\Theta (x)
\end{eqnarray}
at level $(2,7)$, and 
\begin{eqnarray}
&&\partial ^{5}\bar{\partial}^{5}T\bar{T}(x)=\partial ^{5}\bar{\partial}%
^{5}\left( T\bar{T}\right) (x)  \label{Q5-Q5bar-ttbar} \\
&&Q_{5}\bar{Q}_{5}T\bar{T}(x)=Q_{5}\bar{Q}_{5}\left( T\bar{T}\right) (x)-%
\frac{5}{7}R_{6}\bar{\partial}\bar{Q}_{5}\Theta (x)-\frac{5}{7}\partial Q_{5}%
\bar{R}_{6}\Theta (x) \\
&&\partial ^{5}\bar{Q}_{5}T\bar{T}(x)=\partial ^{5}\bar{Q}_{5}\left( T\bar{T}%
\right) (x)-\frac{5}{7}\partial ^{6}\bar{R}_{6}\Theta (x) \\
&&\bar{\partial}^{5}Q_{5}T\bar{T}(x)=\bar{\partial}^{5}Q_{5}\left( T\bar{T}%
\right) (x)-\frac{5}{7}R_{6}\bar{\partial}^{6}\Theta (x)  \label{qt5}
\end{eqnarray}
at level $(7,7)$.

We are now in the position of dealing straightforwardly with the operators $%
S_{4}T\bar{T}$, $\bar{S}_{4}T\bar{T}$ and $S_{4}\bar{S}_{4}T\bar{T}$.
Indeed, the construction above ensures that the equations 
\begin{equation}
Q_{5}T\bar{T}=\partial S_{4}T\bar{T}\hspace{0.1in};\hspace{0.1in}\bar{Q}_{5}T%
\bar{T}=\bar{\partial}\bar{S}_{4}T\bar{T}\hspace{0.1in};\hspace{0.1in}Q_{5}%
\bar{Q}_{5}T\bar{T}=\partial \bar{\partial}S_{4}\bar{S}_{4}T\bar{T}
\label{s4}
\end{equation}
hold also away from criticality, so that we can write the form factors 
\begin{align}
& F_{n}^{S_{4}T\bar{T}}=\left( -im\sigma _{1}^{(n)}\right) ^{-1}\left(
\Lambda _{n}^{\left( 5\right) }F_{n}^{T\bar{T}}+\frac{5}{7}\frac{\bar{\sigma}%
_{1}^{(n)}}{\sigma _{1}^{(n)}}\Lambda _{n}^{\left( 7\right) }F_{n}^{\Theta
}\right) \\
&  \notag \\
& F_{n}^{\bar{S}_{4}T\bar{T}}=\left( im\bar{\sigma}_{1}^{(n)}\right)
^{-1}\left( \Lambda _{n}^{\left( -5\right) }F_{n}^{T\bar{T}}+\frac{5}{7}%
\frac{\sigma _{1}^{(n)}}{\bar{\sigma}_{1}^{(n)}}\Lambda _{n}^{\left(
-7\right) }F_{n}^{\Theta }\right) \\
&  \notag \\
& F_{n}^{S_{4}\bar{S}_{4}T\bar{T}}=\left( m^{2}\sigma _{1}^{(n)}\bar{\sigma}%
_{1}^{(n)}\right) ^{-1}\left( \Lambda _{n}^{\left( 5\right) }\Lambda
_{n}^{\left( -5\right) }F_{n}^{T\bar{T}}+\frac{5}{7}\frac{\bar{\sigma}%
_{1}^{(n)}}{\sigma _{1}^{(n)}}\Lambda _{n}^{\left( 7\right) }\Lambda
_{n}^{\left( -5\right) }F_{n}^{\Theta }+\frac{5}{7}\frac{\sigma _{1}^{(n)}}{%
\bar{\sigma}_{1}^{(n)}}\Lambda _{n}^{\left( -7\right) }\Lambda _{n}^{\left(
5\right) }F_{n}^{\Theta }\right) \,.
\end{align}

Since all the non-chiral descendants of the identity up to level 7 can be
obtained taking derivatives of the operators $T\bar{T}$, $S_{4}T\bar{T}$, $%
\bar{S}_{4}T\bar{T}$ and $S_{4}\bar{S}_{4}T\bar{T}$, our discussion is now
complete.

The factorization properties of the eigenvalues of the conserved quantities
together with (\ref{T-descendant}), (\ref{Tbar-descendant}), (\ref
{clustertheta}) and 
\begin{equation}
\lim_{\alpha \rightarrow +\infty }e^{-2\alpha }F_{n}^{T\bar{T}}\left( \theta
_{1}+\alpha ,..,\theta _{k}+\alpha ,\theta _{k+1},..,\theta _{n}\right)
=F_{k}^{T}\left( \theta _{1},..,\theta _{k}\right) F_{n-k}^{\bar{T}}\left(
\theta _{k+1},..,\theta _{n}\right)  \label{clusterttbar}
\end{equation}
($k=0,1,\ldots ,n$) imply that the form factors of the descendants of the
identity we determined satisfy (\ref{Conjecture-2}) with $\Phi _{0}=I$. In
particular, the chiral and non-chiral descendant satisfy the asymptotic
behavior (\ref{chiral-descendants-Identity}) and (\ref{level-descendant}),
respectively.

\resection{Comparison with the form factor bootstrap}

In the previous section we used the conserved quantities to construct,
starting from the form factor solutions for the energy-momentum tensor and $T%
\bar{T}$, all the operators in the massive Lee-Yang model as expected from
the conformal structure of the operator space, level by level up to level 7.
In this section we want to show that exactly the same operator space is
generated by the solutions of the form factor bootstrap equations (\ref{fn0}%
)-(\ref{fn4}) supplemented by the level-dependent specification (\ref
{level-descendant}) on the asymptotic behavior. We do this explicitly for
the scalar operators of level $(l,l)$ up to $l=7$, thus generalizing what
had been done in \cite{ttbar} for $l\leq 2$.

An essential role in this analysis is played by the so-called kernel
solutions. We call minimal scalar $N$-particle kernel the solution 
\begin{equation}
F_{N}^{K_{N}}(\theta _{1},...,\theta _{N})=2^{N(N-1)}H_{N}\prod_{1\leq
i<j\leq N}F_{min}(\theta _{i}-\theta _{j})  \label{N-kernel}
\end{equation}
of the form factor equations (\ref{fn0})--(\ref{fn4}) with $s_\Phi=0$, $n=N$
and zero on the r.h.s. of (\ref{fn2}) and (\ref{fn4}). This kernel is
minimal in the sense that it has the mildest asymptotic behavior for large
rapidities. The other scalar $N$-particle kernels are given by 
\begin{equation}
U_{N}\left( \theta _{1},..,\theta _{N}\right) F_{N}^{K_{N}}(\theta
_{1},...,\theta _{N})\,,  \label{N-kernels}
\end{equation}
where $U_{N}\left( \theta _{1},..,\theta _{N}\right) $\ is a scalar entire
function of the rapidities, symmetric and (up to a factor $(-1)^{N-1}$) $%
2\pi i$-periodic in all $\theta _{j}$'s. We call $N$-particle kernel
solution the solution $F^{U_NF_N}_n$ of the form factor equations having (%
\ref{N-kernels}) as initial condition ($F^{U_NF_N}_n=0$ for $n<N$ and the $M$%
-particle kernel solutions arising for $M>N$ are ignored).

It follows from (\ref{asyp-fmin}) that 
\begin{equation}
F_{N}^{K_{N}}\left( \theta _{1}+\alpha ,..,\theta _{k}+\alpha ,\theta
_{k+1},...,\theta _{n}\right) \sim e^{k(N-k)\alpha },
\label{asymptotic-N-kernel}
\end{equation}
for $\alpha\rightarrow +\infty$, $N>1$ and $1\leq k\leq N-1$. So, the
maximal asymptotic behavior of $F_{N}^{K_{N}}$ is $e^{d_{K_{N}}\alpha }$
with $d_{K_{2M}}=M^{2}$ and $d_{K_{2M+1}}=(M+1)M$. Similarly, if $%
e^{d_{U_NK_N}}$ is the maximal asymptotic behavior of (\ref{N-kernels}), we
know that $d_{U_NK_N}\geq d_{K_N}$. On the other hand, it is implied by $%
\left( \ref{level-descendant}\right) $ that $U_{N}F_{N}^{K_{N}}$ can
contribute to the form factor solutions of operators of level $(l,l)$ if and
only if $d_{U_{N}K_{N}}=l$. Since $d_{K_N}>7$ for $N>5$, our analysis for $%
l\leq 7$ can involve only $N$-particle kernel solutions with $N\leq 5$.

In table~3 we list, for each level up to $(7,7)$, all the independent scalar
solutions of the form factor equations supplemented by the asymptotic
condition (\ref{level-descendant}). They are given in terms of the form
factors of $\Theta$ and of the kernel solutions labeled as specified in
table~4.

\begin{table}[tbp]
\begin{center}
\begin{tabular}{|l|l|l|l|l|l|l|l|}
\hline
$(0,0)$ & $(1,1)$ & $(2,2)$ & $(3,3)$ & $(4,4)$ & $(5,5)$ & $(6,6)$ & $(7,7)$
\\ \hline\hline
$F_{n}^{\Theta }$ & $D_{n}F_{n}^{\Theta }$ & $\left( D_{n}\right)
^{2}F_{n}^{\Theta }$ & $\left( D_{n}\right) ^{3}F_{n}^{\Theta }$ & $\left(
D_{n}\right) ^{4}F_{n}^{\Theta }$ & $\left( D_{n}\right) ^{5}F_{n}^{\Theta }$
& $\left( D_{n}\right) ^{6}F_{n}^{\Theta }$ & $\left( D_{n}\right)
^{7}F_{n}^{\Theta }$ \\ \hline
$F_{n}^{I}=\delta _{n,0}$ &  & $F_{n}^{K_{3}}$ & $D_{n}F_{n}^{K_{3}}$ & $%
\left( D_{n}\right) ^{2}F_{n}^{K_{3}}$ & $\left( D_{n}\right)
^{3}F_{n}^{K_{3}}$ & $\left( D_{n}\right) ^{4}F_{n}^{K_{3}}$ & $\left(
D_{n}\right) ^{5}F_{n}^{K_{3}}$ \\ \hline
&  &  &  & $F_{n}^{A,K_{3}}$ & $D_{n}F_{n}^{A,K_{3}}$ & $\left( D_{n}\right)
^{2}F_{n}^{A,K_{3}}$ & $\left( D_{n}\right) ^{3}F_{n}^{A,K_{3}}$ \\ \hline
&  &  &  & $F_{n}^{B,K_{3}}$ & $D_{n}F_{n}^{B,K_{3}}$ & $\left( D_{n}\right)
^{2}F_{n}^{B,K_{3}}$ & $\left( D_{n}\right) ^{3}F_{n}^{B,K_{3}}$ \\ \hline
&  &  &  &  &  & $F_{n}^{C,K_{3}}$ & $D_{n}F_{n}^{C,K_{3}}$ \\ \hline
&  &  &  &  &  & $F_{n}^{D,K_{3}}$ & $D_{n}F_{n}^{D,K_{3}}$ \\ \hline
&  &  &  & $F_{n}^{K_{4}}$ & $D_{n}F_{n}^{K_{4}}$ & $\left( D_{n}\right)
^{2}F_{n}^{K_{4}}$ & $\left( D_{n}\right) ^{3}F_{n}^{K_{4}}$ \\ \hline
&  &  &  &  &  & $F_{n}^{A,K_{4}}$ & $D_{n}F_{n}^{A,K_{4}}$ \\ \hline
&  &  &  &  &  & $F_{n}^{B,K_{4}}$ & $D_{n}F_{n}^{B,K_{4}}$ \\ \hline
&  &  &  &  &  & $F_{n}^{C,K_{4}}$ & $D_{n}F_{n}^{C,K_{4}}$ \\ \hline
&  &  &  &  &  & $F_{n}^{D,K_{4}}$ & $D_{n}F_{n}^{D,K_{4}}$ \\ \hline
&  &  &  &  &  & $F_{n}^{E,K_{4}}$ & $D_{n}F_{n}^{E,K_{4}}$ \\ \hline
&  &  &  &  &  & $F_{n}^{K_{5}}$ & $D_{n}F_{n}^{K_{5}}$ \\ \hline\hline
2 & 1 & 2 & 2 & 5 & 5 & 13 & 13 \\ \hline
\end{tabular}
\end{center}
\caption{Scalar solutions of the form factor equations. The first line
specifies the level, the last line the number of solutions. $D_{n}\equiv
\left( \protect\sigma _{1}^{(n)}\bar{\protect\sigma}_{1}^{(n)}\right) $ is $%
m^{-2}$ times the eigenvalue of $\partial \bar{\partial}$ on the $n$%
-particle asymptotic state.}
\end{table}

\begin{table}[tbp]
\begin{center}
\begin{tabular}{|l|l|}
\hline
Kernel solution & Initial condition \\ \hline\hline
$F_{n}^{K_N}$ & $F_{N}^{K_{N}}$ \\ \hline
$F_{n}^{A,K_{3}}$ & $\left( \sigma _{1}^{(3)}\right) ^{2}\bar{\sigma}%
_{2}^{(3)}F_{3}^{K_{3}}$ \\ \hline
$F_{n}^{B,K_{3}}$ & $\sigma _{2}^{(3)}\left( \bar{\sigma}_{1}^{(3)}\right)
^{2}F_{3}^{K_{3}}$ \\ \hline
$F_{n}^{C,K_{3}}$ & $\left( \sigma _{1}^{(3)}\right) ^{4}\left( \bar{\sigma}%
_{2}^{(3)}\right) ^{2}F_{3}^{K_{3}}$ \\ \hline
$F_{n}^{D,K_{3}}$ & $\left( \sigma _{2}^{(3)}\right) ^{2}\left( \bar{\sigma}%
_{1}^{(3)}\right) ^{4}F_{3}^{K_{3}}$ \\ \hline
$F_{n}^{A,K_{4}}$ & $\left( \sigma _{1}^{(4)}\right) ^{2}\bar{\sigma}%
_{2}^{(4)}F_{4}^{K_{4}}$ \\ \hline
$F_{n}^{B,K_{4}}$ & $\sigma _{2}^{(4)}\left( \bar{\sigma}_{1}^{(4)}\right)
^{2}F_{4}^{K_{4}}$ \\ \hline
$F_{n}^{C,K_{4}}$ & $\sigma _{2}^{(4)}\bar{\sigma}_{2}^{(4)}F_{4}^{K_{4}}$
\\ \hline
$F_{n}^{D,K_{4}}$ & $\left( \sigma _{1}^{(4)}\right) ^{3}\bar{\sigma}%
_{3}^{(4)}F_{4}^{K_{4}}$ \\ \hline
$F_{n}^{E,K_{4}}$ & $\sigma _{3}^{(4)}\left( \bar{\sigma}_{1}^{(4)}\right)
^{3}F_{4}^{K_{4}}$ \\ \hline
\end{tabular}
\end{center}
\caption{{The first column lists the names given to the form factor kernel
solutions originating from the initial conditions specified in the second
column. }}
\end{table}

The asymptotic properties 
\begin{align*}
\lim_{\alpha \rightarrow +\infty }e^{-k\alpha }\sigma _{p}^{(n)}\left(
x_{1}e^{\alpha },..,x_{k}e^{\alpha },x_{k+1},..,x_{n}\right) & =\sigma
_{k}^{(k)}(x_{1},\ldots ,x_{k})\sigma _{p-k}^{(n-k)}(x_{k+1},\ldots ,x_{n}),%
\hspace{0.1in}k\leq p \\
& \\
\lim_{\alpha \rightarrow +\infty }e^{-p\alpha }\sigma _{p}^{(n)}\left(
x_{1}e^{\alpha },..,x_{k}e^{\alpha },x_{k+1},..,x_{n}\right) & =\sigma
_{p}^{(k)}(x_{1},\ldots ,x_{k}),\hspace{0.1in}\hspace{0.1in}\hspace{0.1in}%
k\geq p
\end{align*}
of the elementary symmetric polynomials are useful to check that each kernel
solution corresponds to the level indicated in table~3.

Tables~5 and 6 list the scalar descendants of $\Theta$ and $I$,
respectively, up to level $(7,7)$, as constructed in the previous section by
continuing away from criticality the conformal basis. Notice that, for each
level, the total number of independent operators obtained putting together
the two operator families perfectly matches the number of solutions of the
form factor boostrap for that level given in table~3. Hence, the last thing
to check in order to show that the solutions of table~3 span the same space
of those of tables~5 and 6 is that the kernel solutions of table~4 can be
rewritten as linear combinations of the solutions of tables~5 and 6. This
change of basis is given explicitly in Appendix~C.

\begin{table}[tbp]
\begin{center}
\begin{tabular}{|l|l|l|l|l|l|l|l|}
\hline
$(0,0)$ & $(1,1)$ & $(2,2)$ & $(3,3)$ & $(4,4)$ & $(5,5)$ & $(6,6)$ & $(7,7)$
\\ \hline\hline
$\overset{}{F_{n}^{\Theta }}$ & $\overset{}{F_{n}^{\partial \bar{\partial}%
\Theta }}$ & $\overset{}{F_{n}^{\partial ^{2}\bar{\partial}^{2}\Theta }}$ & $%
\overset{}{F_{n}^{\partial ^{3}\bar{\partial}^{3}\Theta }}$ & $\overset{}{%
F_{n}^{\partial ^{4}\bar{\partial}^{4}\Theta }}$ & $\overset{}{%
F_{n}^{\partial ^{5}\bar{\partial}^{5}\Theta }}$ & $\overset{}{%
F_{n}^{\partial ^{6}\bar{\partial}^{6}\Theta }}$ & $\overset{}{%
F_{n}^{\partial ^{7}\bar{\partial}^{7}\Theta }}$ \\ \hline
$\overset{}{}$ & $\overset{}{}$ & $\overset{}{}$ & $\overset{}{}$ & $%
\overset{}{F_{n}^{R_{4}\bar{R}_{4}\Theta }}$ & $\overset{}{F_{n}^{Q_{5}\bar{Q%
}_{5}\Theta }}$ & $\overset{}{F_{n}^{\partial Q_{5}\bar{\partial}\bar{Q}%
_{5}\Theta }}$ & $\overset{}{F_{n}^{\partial ^{2}Q_{5}\bar{\partial}^{2}\bar{%
Q}_{5}\Theta }}$ \\ \hline
$\overset{}{}$ & $\overset{}{}$ & $\overset{}{}$ & $\overset{}{}$ & $%
\overset{}{F_{n}^{\partial ^{4}\bar{R}_{4}\Theta }}$ & $\overset{}{%
F_{n}^{\partial ^{5}\bar{Q}_{5}\Theta }}$ & $\overset{}{F_{n}^{\partial ^{6}%
\bar{\partial}\bar{Q}_{5}\Theta }}$ & $\overset{}{F_{n}^{\partial ^{7}\bar{%
\partial}^{2}\bar{Q}_{5}\Theta }}$ \\ \hline
$\overset{}{}$ & $\overset{}{}$ & $\overset{}{}$ & $\overset{}{}$ & $%
\overset{}{F_{n}^{R_{4}\bar{\partial}^{4}\Theta }}$ & $\overset{}{%
F_{n}^{Q_{5}\bar{\partial}^{5}\Theta }}$ & $\overset{}{F_{n}^{\partial Q_{5}%
\bar{\partial}^{6}\Theta }}$ & $\overset{}{F_{n}^{\partial ^{2}Q_{5}\bar{%
\partial}^{7}\Theta }}$ \\ \hline
$\overset{}{}$ & $\overset{}{}$ & $\overset{}{}$ & $\overset{}{}$ & $%
\overset{}{}$ & $\overset{}{}$ & $\overset{}{F_{n}^{R_{6}\bar{\partial}%
^{6}\Theta }}$ & $\overset{}{F_{n}^{Q_{7}\bar{\partial}^{7}\Theta }}$ \\ 
\hline
$\overset{}{}$ & $\overset{}{}$ & $\overset{}{}$ & $\overset{}{}$ & $%
\overset{}{}$ & $\overset{}{}$ & $\overset{}{F_{n}^{\partial ^{6}\bar{R}%
_{6}\Theta }}$ & $\overset{}{F_{n}^{\partial ^{7}\bar{Q}_{7}\Theta }}$ \\ 
\hline
$\overset{}{}$ & $\overset{}{}$ & $\overset{}{}$ & $\overset{}{}$ & $%
\overset{}{}$ & $\overset{}{}$ & $\overset{}{F_{n}^{R_{6}\bar{R}_{6}\Theta }}
$ & $\overset{}{F_{n}^{Q_{7}\bar{Q}_{7}\Theta }}$ \\ \hline
$\overset{}{}$ & $\overset{}{}$ & $\overset{}{}$ & $\overset{}{}$ & $%
\overset{}{}$ & $\overset{}{}$ & $\overset{}{F_{n}^{\partial Q_{5}\bar{R}%
_{6}\Theta }}$ & $\overset{}{F_{n}^{\partial ^{2}Q_{5}\bar{Q}_{7}\Theta }}$
\\ \hline
$\overset{}{}$ & $\overset{}{}$ & $\overset{}{}$ & $\overset{}{}$ & $%
\overset{}{}$ & $\overset{}{}$ & $\overset{}{F_{n}^{R_{6}\bar{\partial}\bar{Q%
}_{5}\Theta }}$ & $\overset{}{F_{n}^{Q_{7}\bar{\partial}^{2}\bar{Q}%
_{5}\Theta }}$ \\ \hline\hline
1 & 1 & 1 & 1 & 4 & 4 & 9 & 9 \\ \hline
\end{tabular}
\end{center}
\caption{Scalar operators in the family of $\Theta $ up to level $(7,7)$.
The first and last line are as in table~3.}
\end{table}

\begin{table}[tbp]
\begin{center}
\begin{tabular}[b]{|c|c|c|c|c|c|c|c|}
\hline
$(0,0)$ & $(1,1)$ & $(2,2)$ & $(3,3)$ & $(4,4)$ & $(5,5)$ & $(6,6)$ & $(7,7)$
\\ \hline\hline
$\overset{}{F_{n}^{I}=\delta _{n,0}}$ & $\overset{}{}$ & $\overset{}{{F}%
_{n}^{T\bar{T}}}$ & $\overset{}{F_{n}^{\partial \bar{\partial}T\bar{T}}}$ & $%
\overset{}{F_{n}^{\partial ^{2}\bar{\partial}^{2}T\bar{T}}}$ & $\overset{}{%
F_{n}^{\partial ^{3}\bar{\partial}^{3}T\bar{T}}}$ & $\overset{}{%
F_{n}^{\partial ^{4}\bar{\partial}^{4}T\bar{T}}}$ & $\overset{}{%
F_{n}^{\partial ^{5}\bar{\partial}^{5}T\bar{T}}}$ \\ \hline
&  &  &  &  &  & $\overset{}{F_{n}^{S_{4}\bar{S}_{4}T\bar{T}}}$ & $\overset{%
}{F\vspace{0.02in}_{n}^{Q_{5}\bar{Q}_{5}T\bar{T}}}$ \\ \hline
&  &  &  &  &  & $\overset{}{F_{n}^{\partial ^{4}\bar{S}_{4}T\bar{T}}}$ & $%
\overset{}{F_{n}^{\partial ^{5}\bar{Q}_{5}T\bar{T}}}$ \\ \hline
&  &  &  &  &  & $\overset{}{F_{n}^{S_{4}\bar{\partial}^{4}T\bar{T}}} $ & $%
\overset{}{F_{n}^{Q_{5}\bar{\partial}^{5}T\bar{T}}}$ \\ \hline\hline
1 & 0 & 1 & 1 & 1 & 1 & 4 & 4 \\ \hline
\end{tabular}
\end{center}
\caption{Scalar operators in the family of $I$ up to level $(7,7)$. The
first and last line are as in table~3.}
\end{table}

\resection{Conclusion}

This paper provides the first extensive study of the operator space of a
massive two-dimensional quantum field theory aimed at the explicit
construction of descendant operators. From the point of view of the $S$%
-matrix bootstrap, in particular, we have shown how the identification of
operator subspaces can be realized supplementing the form factor equations
with asymptotic conditions containing the information about the level of the
descendants, thus extending what originally done in \cite{immf,DS,DSC} for
the primary operators.

The space of solutions generated in this way has been shown to coincide with
the off-critical continuation of the conformal operator space obtained
acting with the conserved quantities of the massive Lee-Yang model on the
lowest non-trivial form factor solutions in the two operator families. The
fact that not all the operators beyond level 7 can be generated using the
conserved quantities is the model dependent feature that has determined the
extent of our analysis of the operator space in this paper.

In constructing the composite descendant operators in the operator family of
the identity we found that the proliferation of undetermined coefficients of
resonant operators expected on purely dimensional grounds is actually
limited by the structure of the operator space. More precisely, the
requirement that some composite operators remain derivative operators also
away from criticality can indeed be fulfilled by fixing in a unique way the
coefficients of some resonant terms, with the consequence that only the
"primitive" ambiguity contained in the definition of $T\bar{T} $ propagates
through the operator family up to level 7. This as well as the previous
points deserve to be further investigated in other models.

\vspace{1cm} \textbf{Acknowledgments.}~~This work was partially supported by
the European Commission programme HPRN-CT-2002-00325 (EUCLID) and by the
COFIN ``Teoria dei Campi, Meccanica Statistica e Sistemi Elettronici''.

\appendix

\resection{Appendix}

Let $Q_{s}\bar{Q}_{r}T\bar{T}$ be the off-critical continuation of the
descendant $Q_{s}\bar{Q}_{r}L_{-2}\bar{L}_{-2}I$ obtained through the
regularization 
\begin{equation}
Q_{s}\bar{Q}_{r}T\bar{T}(x)=\lim_{x^{\prime }\rightarrow x}\left\{ \left[
Q_{s},\left[ \bar{Q}_{r},T(x)\bar{T}(x^{\prime })\right] \right] -\mathcal{R}%
_{Q_{s}\bar{Q}_{r}T\bar{T}}(x,x^{\prime })\right\} ,  \label{q-qbar-ttbar}
\end{equation}
with $\mathcal{R}_{Q_{s}\bar{Q}_{r}T\bar{T}}(x,x^{\prime })$ such that the
above limit is non-singular. Then the matrix elements of $\left( \ref
{q-qbar-ttbar}\right) $ 
\begin{eqnarray}
&&\langle \theta _{m}^{\prime }\ldots \theta _{1}^{\prime }|Q_{s}\bar{Q}_{r}T%
\bar{T}(x)|\theta _{1}\ldots \theta _{n}\rangle =  \notag \\
&=&\lim_{x^{\prime }\rightarrow x}\langle \theta _{m}^{\prime }\ldots \theta
_{1}^{\prime }|\left\{ \left[ Q_{s},\left[ \bar{Q}_{r},T(x)\bar{T}(x^{\prime
})\right] \right] -\mathcal{R}_{Q_{s}\bar{Q}_{r}T\bar{T}}(x,x^{\prime
})\right\} |\theta _{1}\ldots \theta _{n}\rangle
\end{eqnarray}
must be finite for any $x$. Since 
\begin{align}
& \left. \langle \theta _{m}^{\prime }\ldots \theta _{1}^{\prime }|\left[
Q_{s},\left[ \bar{Q}_{r},T(x)\bar{T}(x^{\prime })\right] \right] |\theta
_{1}\ldots \theta _{n}\rangle =\left( \Lambda _{m}^{\left( s\right) }\left(
\theta _{1}^{\prime },..,\theta _{m}^{\prime }\right) -\Lambda _{n}^{\left(
s\right) }\left( \theta _{1},..,\theta _{n}\right) \right) \times \right. 
\notag \\
& \,\text{\ \ \ \ \ \ \ \ \ \ \ \ \ \ \ \ \ \ \ \ \ \ \ \ \ \ \ \ \ \ \ \ }%
\times \left( \Lambda _{m}^{\left( -r\right) }\left( \theta _{1}^{\prime
},..,\theta _{m}^{\prime }\right) -\Lambda _{n}^{\left( -r\right) }\left(
\theta _{1},..,\theta _{n}\right) \right) \langle \theta _{m}^{\prime
}\ldots \theta _{1}^{\prime }|T(x)\bar{T}(x^{\prime })|\theta _{1}\ldots
\theta _{n}\rangle
\end{align}
the regularization term is fixed to 
\begin{equation}
\mathcal{R}_{Q_{s}\bar{Q}_{r}T\bar{T}}(x,x^{\prime })=\left[ Q_{s},\left[ 
\bar{Q}_{r},\mathcal{R}_{T\bar{T}}(x,x^{\prime })\right] \right] +\text{%
resonances}(x)\,,
\end{equation}
where $\mathcal{R}_{T\bar{T}}(x,x^{\prime })$ is the regularization term (%
\ref{regttbar}) of the operator $T\bar{T}$. Putting all together we have 
\begin{equation}
Q_{s}\bar{Q}_{r}T\bar{T}(x)=Q_{s}\bar{Q}_{r}\left( T\bar{T}\right) (x)+\text{%
resonances}(x),
\end{equation}
where $Q_{s}\bar{Q}_{r}\left( T\bar{T}\right) $ is the local field obtained
acting with the conserved commuting quantities $Q_{s}\bar{Q}_{r}$ on the
operator (\ref{regularized}).

\resection{Appendix}

Consider a massive two-dimensional field theory without internal symmetries.
It was argued in \cite{DSC} that in such a case the form factors of a
primary operator $\Phi _{0}$ satisfy the asymptotic factorization property 
\begin{gather}
\lim_{\alpha \longrightarrow +\infty }\text{ }F_{n}^{\Phi _{0}}\left( \theta
_{1}+\alpha ,..,\theta _{k}+\alpha ,\theta _{k+1},..,\theta _{n}\right) =%
\hspace{0.4in}\hspace{0.15in}\hspace{0.15in}\hspace{0.15in}\hspace{0.15in}%
\hspace{0.15in}  \notag \\
\hspace{0.15in}\hspace{0.15in}\hspace{0.15in}\hspace{0.15in}\hspace{0.4in}%
\hspace{0.4in}=\frac{1}{\left\langle \Phi _{0}\right\rangle }F_{k}^{\Phi
_{0}}\left( \theta _{1},..,\theta _{k}\right) F_{n-k}^{\Phi _{0}}\left(
\theta _{k+1},..,\theta _{n}\right) \text{,}  \label{primaryfactorization}
\end{gather}
for $k=0,1,\ldots ,n$. Here we consider the effect of the same limit on the
descendant operators 
\begin{equation*}
\Phi^{\left\{ s_{1},..,s_{a}\right\} \left\{ r_{1},..,r_{b}\right\} }\left(
x\right) =[\bar{Q}_{r_{b}},[...,[\bar{Q}_{r_{1}},[Q_{s_{a}},[...,[Q_{s_{1}},%
\Phi _{0}\left( x\right) ]]...]]...]]
\end{equation*}
obtained acting on $\Phi _{0}$ with the commuting conserved quantities $%
Q_{s} $ and $\bar{Q}_{r}$ ($r$ and $s$ positive integers). The level of such
operators is $(l,\bar{l})=\left(
\sum_{i=1}^{a}s_{i},\sum_{j=1}^{b}r_{j}\right) $.

The form factors for these operators take the form 
\begin{equation}
F_{n}^{\Phi^{\{s_{1},..,s_{a}\}\{r_{1},..,r_{b}\}}}\left( \theta
_{1},..,\theta _{n}\right) =\left( -1\right) ^{l+\bar{l}}\Lambda
_{n}^{\left\{ s_{1},..,s_{a}\right\} \left\{ r_{1},..,r_{b}\right\} }\left(
\theta _{1},..,\theta _{n}\right) F_{n}^{\Phi _{0}}\left( \theta
_{1},..,\theta _{n}\right)  \label{ccc-descendent-form-factors}
\end{equation}
where 
\begin{equation}
\Lambda _{n}^{\left\{ s_{1},..,s_{a}\right\} \left\{ r_{1},..,r_{b}\right\}
}\left( \theta _{1},..,\theta _{n}\right) =\prod_{\mu =1}^{a}\Lambda
_{n}^{\left( s_{\mu }\right) }\left( \theta _{1},..,\theta _{n}\right)
\prod_{\upsilon =1}^{b}\Lambda _{n}^{\left( -r_{\upsilon }\right) }\left(
\theta _{1},..,\theta _{n}\right)  \label{landa-functions-2}
\end{equation}
with $\Lambda _{n}^{\left( s\right) }$ defined in $\left( \ref
{landa-functions}\right) $.

It follows from (\ref{primaryfactorization}) and the asymptotic
factorization properties 
\begin{gather}
\lim_{\alpha \longrightarrow +\infty }\text{ }e^{-s\alpha }\Lambda
_{n}^{\left( s\right) }\left( \theta _{1}+\alpha ,..,\theta _{k}+\alpha
,\theta _{k+1},..,\theta _{n}\right) =\lim_{\alpha \longrightarrow +\infty }%
\text{ }e^{-s\alpha }m^{s}(\sum_{i=1}^{k}e^{s\left( \theta _{i}+\alpha
\right) }+  \notag \\
+\sum_{i=k+1}^{n}e^{s\theta _{i}})=m^{s}\sum_{i=1}^{k}e^{s\theta
_{i}}=\Lambda _{k}^{\left( s\right) }\left( \theta _{1},..,\theta
_{k}\right) ,
\end{gather}
\begin{gather}
\lim_{\alpha \longrightarrow +\infty }\text{ }\Lambda _{n}^{\left( -r\right)
}\left( \theta _{1}+\alpha ,..,\theta _{k}+\alpha ,\theta _{k+1},..,\theta
_{n}\right) =\lim_{\alpha \longrightarrow +\infty }\text{ }%
m^{r}(\sum_{i=1}^{k}e^{-r\left( \theta _{i}+\alpha \right) }+  \notag \\
+\sum_{i=k+1}^{n}e^{-r\theta _{i}})=m^{r}\sum_{i=k+1}^{n}e^{-r\theta
_{i}}=\Lambda _{n-k}^{\left( -r\right) }\left( \theta _{k+1},..,\theta
_{n}\right) ,
\end{gather}
that the form factors of $\Phi^{\left\{ s_{1},..,s_{a}\right\} \left\{
r_{1},..,r_{b}\right\} }$ satisfy 
\begin{gather}
\lim_{\alpha \longrightarrow +\infty }\text{ }e^{-l\alpha
}F_{n}^{\Phi^{\{s_{1},..,s_{a}\}\{r_{1},..,r_{b}\}}}\left( \theta
_{1}+\alpha ,..,\theta _{k}+\alpha ,\theta _{k+1},..,\theta _{n}\right) =%
\hspace{0.4in}\hspace{0.15in}\hspace{0.15in}\hspace{0.15in}\hspace{0.15in}%
\hspace{0.15in}  \notag \\
\hspace{0.15in}\hspace{0.15in}\hspace{0.15in}\hspace{0.15in}\hspace{0.4in}%
\hspace{0.4in}=\frac{1}{\left\langle \Phi _{0}\right\rangle }%
F_{k}^{\Phi^{\{s_{1},..,s_{a}\}\{\}}}\left( \theta _{1},..,\theta
_{k}\right) F_{n-k}^{\Phi^{\{\}\{r_{1},..,r_{b}\}}}\left( \theta
_{k+1},..,\theta _{n}\right)  \label{PROPRIETY-2}
\end{gather}
for $k=0,1,\ldots ,n$. Hence, the form factors of the generic $(l,\bar{l})$
descendant $\Phi^{\left\{ s_{1},..,s_{a}\right\} \left\{
r_{1},..,r_{b}\right\} }$ factorizes into the form factors of the chiral
descendants $\Phi^{\left\{ s_{1},..,s_{a}\right\} \left\{ {}\right\} }$ and $%
\Phi^{\left\{ {}\right\} \left\{ r_{1},..,r_{b}\right\} }$ of level $(l,0)$
and $(0,\bar{l})$, respectively.

Denote by $\Phi _{(l,\bar{l})}$ a generic linear combination of the level $%
(l,\bar{l})$ descendants $\Phi^{\left\{ s_{1},..,s_{a}\right\} \left\{
r_{1},..,r_{b}\right\} }$ of a scalar primary other than the identity. Then (%
\ref{PROPRIETY-2}) implies that the form factors of $\Phi _{(l,\bar{l})}$
are homogeneous functions, with respect to the variables $x_{i}\equiv
e^{\theta _{i}}$, of degree $l-\bar{l}$ with further asymptotic behavior 
\begin{equation}
F_{n}^{\Phi _{(l,\bar{l})}}\left( \theta _{1}+\alpha ,..,\theta _{k}+\alpha
,\theta _{k+1},...,\theta _{n}\right) \sim e^{l\alpha }  \label{PROPRIETY-1}
\end{equation}
for $\alpha\rightarrow +\infty$, $n>1$ and $1\leq k\leq n-1$.

\resection{Appendix}

We give here the expansions of the kernel solutions appearing in table~3 in
terms of the solutions of tables~5 and 6. The kernel solution $F_{n}^{K_{3}}$
arising at level $(2,2)$ is related to $F_{n}^{T\bar{T}}$ by (\ref{fnttbar}%
). Throughout this appendix we take $c=0$ in (\ref{fnttbar}). Then we have

\begin{equation}
\begin{array}{c}
F_{n}^{A,K_{3}}=\frac{1}{5m^{8}\langle \Theta \rangle }\left(
F_{n}^{\partial ^{4}\bar{\partial}^{4}\Theta }-iF_{n}^{\partial ^{4}\bar{R}%
_{4}\Theta }\right) -\frac{1}{m^{6}\langle \Theta \rangle }F_{n}^{\partial
^{3}\bar{\partial}^{3}\Theta }+\frac{1}{m^{4}\langle \Theta \rangle }%
F_{n}^{\partial ^{2}\bar{\partial}^{2}\Theta }
\end{array}
\text{ \ \ \ \ \ \ \ \ \ \ \ \ \ \ \ \ \ \ \ \ \ \ \ \ \ \ \ \ \ \ \ \ \ \ \
\ \ \ \ \ \ \ \ }
\end{equation}

\begin{equation}
\begin{array}{c}
F_{n}^{B,K_{3}}=\frac{1}{5m^{8}\langle \Theta \rangle }\left(
F_{n}^{\partial ^{4}\bar{\partial}^{4}\Theta }+iF_{n}^{R_{4}\bar{\partial}%
^{4}\Theta }\right) -\frac{1}{m^{6}\langle \Theta \rangle }F_{n}^{\partial
^{3}\bar{\partial}^{3}\Theta }+\frac{1}{m^{4}\langle \Theta \rangle }%
F_{n}^{\partial ^{2}\bar{\partial}^{2}\Theta }
\end{array}
\text{ \ \ \ \ \ \ \ \ \ \ \ \ \ \ \ \ \ \ \ \ \ \ \ \ \ \ \ \ \ \ \ \ \ \ \
\ \ \ \ \ \ \ \ }
\end{equation}
\vspace{0.07in}\vspace{0.07in} 
\begin{equation}
\begin{array}{l}
F_{n}^{K_{4}}=\frac{1}{25m^{8}\langle \Theta \rangle }\left( F_{n}^{\partial
^{4}\bar{\partial}^{4}\Theta }-iF_{n}^{\partial ^{4}\bar{R}_{4}\Theta
}+iF_{n}^{R_{4}\bar{\partial}^{4}\Theta }+F_{n}^{R_{4}\bar{R}_{4}\Theta
}\right) +\frac{1}{m^{6}\langle \Theta \rangle }F_{n}^{\partial ^{3}\bar{%
\partial}^{3}\Theta }-\frac{2}{m^{4}\langle \Theta \rangle }F_{n}^{\partial
^{2}\bar{\partial}^{2}\Theta }+\frac{1}{\langle \Theta \rangle }%
F_{n}^{\Theta } \\ 
\\ 
-\frac{1}{m^{2}\langle \Theta \rangle ^{2}}F_{n}^{\partial \bar{\partial}T%
\bar{T}}+\frac{1}{\langle \Theta \rangle ^{2}}F_{n}^{T\bar{T}}
\end{array}
\text{\ \ \ }
\end{equation}
\begin{equation*}
\end{equation*}
\vspace{0.07in}for the kernels arising at level $(4,4)$, and\vspace{0.07in} 
\begin{equation}
\begin{array}{l}
F_{n}^{C,K_{3}}=\frac{1}{175m^{12}\langle \Theta \rangle }\left(
8F_{n}^{\partial ^{6}\bar{\partial}^{6}\Theta }-5F_{n}^{\partial Q_{5}\bar{R}%
_{6}\Theta }-7F_{n}^{\partial Q_{5}\bar{\partial}\bar{Q}_{5}\Theta
}-28iF_{n}^{\partial ^{6}\bar{\partial}\bar{Q}_{5}\Theta
}-20iF_{n}^{\partial ^{6}\bar{R}_{6}\Theta }-2iF_{n}^{\partial Q_{5}\bar{%
\partial}^{6}\Theta }\right)  \\ 
\\ 
+\frac{1}{5m^{8}\langle \Theta \rangle }\left( -6F_{n}^{\partial ^{4}\bar{%
\partial}^{4}\Theta }+iF_{n}^{\partial ^{4}\bar{R}_{4}\Theta }\right) +\frac{%
3}{m^{6}\langle \Theta \rangle }F_{n}^{\partial ^{3}\bar{\partial}^{3}\Theta
}-\frac{3}{m^{4}\langle \Theta \rangle }F_{n}^{\partial ^{2}\bar{\partial}%
^{2}\Theta }+\frac{1}{m^{2}\langle \Theta \rangle }F_{n}^{\partial \bar{%
\partial}\Theta }\vspace{0.07in}
\end{array}
\end{equation}
\vspace{0.07in}\vspace{0.07in}\vspace{0.07in}\vspace{0.07in} 
\begin{equation}
\begin{array}{l}
F_{n}^{D,K_{3}}\text{ }=\frac{1}{175m^{12}\langle \Theta \rangle }\left(
8F_{n}^{\partial ^{6}\bar{\partial}^{6}\Theta }-5F_{n}^{R_{6}\bar{\partial}%
\bar{Q}_{5}\Theta }-7F_{n}^{\partial Q_{5}\bar{\partial}\bar{Q}_{5}\Theta
}+2iF_{n}^{\partial ^{6}\bar{\partial}\bar{Q}_{5}\Theta }+28iF_{n}^{\partial
Q_{5}\bar{\partial}^{6}\Theta }+20iF_{n}^{R_{6}\bar{\partial}^{6}\Theta
}\right)  \\ 
\\ 
+\frac{1}{5m^{8}\langle \Theta \rangle }\left( -6F_{n}^{\partial ^{4}\bar{%
\partial}^{4}\Theta }-iF_{n}^{R_{4}\bar{\partial}^{4}\Theta }\right) +\frac{3%
}{m^{6}\langle \Theta \rangle }F_{n}^{\partial ^{3}\bar{\partial}^{3}\Theta
}-\frac{3}{m^{4}\langle \Theta \rangle }F_{n}^{\partial ^{2}\bar{\partial}%
^{2}\Theta }+\frac{1}{m^{2}\langle \Theta \rangle }F_{n}^{\partial \bar{%
\partial}\Theta }
\end{array}
\end{equation}
\vspace{0.07in}\vspace{0.07in}\vspace{0.07in}\vspace{0.07in} 
\begin{equation}
\begin{array}{l}
F_{n}^{A,K_{4}}=\frac{1}{175m^{12}\langle \Theta \rangle }\left(
2F_{n}^{\partial ^{6}\bar{\partial}^{6}\Theta }+5F_{n}^{\partial Q_{5}\bar{R}%
_{6}\Theta }+7F_{n}^{\partial Q_{5}\bar{\partial}\bar{Q}_{5}\Theta
}-7iF_{n}^{\partial ^{6}\bar{\partial}\bar{Q}_{5}\Theta }-5iF_{n}^{\partial
^{6}\bar{R}_{6}\Theta }+2iF_{n}^{\partial Q_{5}\bar{\partial}^{6}\Theta
}\right)  \\ 
\\ 
+\frac{1}{5m^{10}\langle \Theta \rangle }\left( -F_{n}^{\partial ^{5}\bar{%
\partial}^{5}\Theta }+iF_{n}^{\partial ^{5}\bar{Q}_{5}\Theta }\right) +\frac{%
1}{5m^{8}\langle \Theta \rangle }\left( 6F_{n}^{\partial ^{4}\bar{\partial}%
^{4}\Theta }-iF_{n}^{\partial ^{4}\bar{R}_{4}\Theta }\right) -\frac{3}{%
m^{6}\langle \Theta \rangle }F_{n}^{\partial ^{3}\bar{\partial}^{3}\Theta }+%
\frac{3}{m^{4}\langle \Theta \rangle }F_{n}^{\partial ^{2}\bar{\partial}%
^{2}\Theta } \\ 
\\ 
-\frac{1}{m^{2}\langle \Theta \rangle }F_{n}^{\partial \bar{\partial}\Theta }
\end{array}
\end{equation}
\vspace{0.07in}\vspace{0.07in}\vspace{0.07in}\vspace{0.07in} 
\begin{equation}
\begin{array}{l}
F_{n}^{B,K_{4}}\text{ }=\frac{1}{175m^{12}\langle \Theta \rangle }\left(
2F_{n}^{\partial ^{6}\bar{\partial}^{6}\Theta }+5F_{n}^{R_{6}\bar{\partial}%
\bar{Q}_{5}\Theta }+7F_{n}^{\partial Q_{5}\bar{\partial}\bar{Q}_{5}\Theta
}-2iF_{n}^{\partial ^{6}\bar{\partial}\bar{Q}_{5}\Theta }+7iF_{n}^{\partial
Q_{5}\bar{\partial}^{6}\Theta }+5iF_{n}^{R_{6}\bar{\partial}^{6}\Theta
}\right)  \\ 
\\ 
+\frac{1}{5m^{10}\langle \Theta \rangle }\left( -F_{n}^{\partial ^{5}\bar{%
\partial}^{5}\Theta }-iF_{n}^{Q_{5}\bar{\partial}^{5}\Theta }\right) +\frac{1%
}{5m^{8}\langle \Theta \rangle }\left( 6F_{n}^{\partial ^{4}\bar{\partial}%
^{4}\Theta }+iF_{n}^{R_{4}\bar{\partial}^{4}\Theta }\right) -\frac{3}{%
m^{6}\langle \Theta \rangle }F_{n}^{\partial ^{3}\bar{\partial}^{3}\Theta }+%
\frac{3}{m^{4}\langle \Theta \rangle }F_{n}^{\partial ^{2}\bar{\partial}%
^{2}\Theta } \\ 
\\ 
-\frac{1}{m^{2}\langle \Theta \rangle }F_{n}^{\partial \bar{\partial}\Theta }
\end{array}
\end{equation}
\vspace{0.07in}\vspace{0.07in}\vspace{0.07in}\vspace{0.07in} 
\begin{equation}
\begin{array}{l}
F_{n}^{C,K_{4}}\text{ }=\frac{1}{1225m^{12}\langle \Theta \rangle }\left(
4F_{n}^{\partial ^{6}\bar{\partial}^{6}\Theta }+25F_{n}^{R_{6}\bar{R}%
_{6}\Theta }+35F_{n}^{R_{6}\bar{\partial}\bar{Q}_{5}\Theta
}+35F_{n}^{\partial Q_{5}\bar{R}_{6}\Theta }+49F_{n}^{\partial Q_{5}\bar{%
\partial}\bar{Q}_{5}\Theta }-14iF_{n}^{\partial ^{6}\bar{\partial}\bar{Q}%
_{5}\Theta }\right.  \\ 
\\ 
\left. -10iF_{n}^{\partial ^{6}\bar{R}_{6}\Theta }+14iF_{n}^{\partial Q_{5}%
\bar{\partial}^{6}\Theta }+10iF_{n}^{R_{6}\bar{\partial}^{6}\Theta }\right) +%
\frac{1}{25m^{10}\langle \Theta \rangle }\left( -F_{n}^{\partial ^{5}\bar{%
\partial}^{5}\Theta }-F_{n}^{Q_{5}\bar{Q}_{5}\Theta }+iF_{n}^{\partial ^{5}%
\bar{Q}_{5}\Theta }-iF_{n}^{Q_{5}\bar{\partial}^{5}\Theta }\right)  \\ 
\\ 
+\frac{1}{25m^{8}\langle \Theta \rangle }\left( F_{n}^{\partial ^{4}\bar{%
\partial}^{4}\Theta }+F_{n}^{R_{4}\bar{R}_{4}\Theta }-iF_{n}^{\partial ^{4}%
\bar{R}_{4}\Theta }+iF_{n}^{R_{4}\bar{\partial}^{4}\Theta }\right) +\frac{1}{%
m^{6}\langle \Theta \rangle }F_{n}^{\partial ^{3}\bar{\partial}^{3}\Theta }-%
\frac{4}{m^{4}\langle \Theta \rangle }F_{n}^{\partial ^{2}\bar{\partial}%
^{2}\Theta }+\frac{5}{m^{2}\langle \Theta \rangle }F_{n}^{\partial \bar{%
\partial}\Theta } \\ 
\\ 
-\frac{2}{\langle \Theta \rangle }F_{n}^{\Theta }
\end{array}
\end{equation}
\vspace{0.07in}\vspace{0.07in}\vspace{0.07in}\vspace{0.07in} 
\begin{equation}
\begin{array}{l}
F_{n}^{D,K_{4}}\text{ }=\frac{1}{245m^{12}\langle \Theta \rangle }\left(
10F_{n}^{\partial ^{6}\bar{\partial}^{6}\Theta }-25F_{n}^{R_{6}\bar{R}%
_{6}\Theta }-35F_{n}^{R_{6}\bar{\partial}\bar{Q}_{5}\Theta
}-35F_{n}^{\partial Q_{5}\bar{R}_{6}\Theta }-84iF_{n}^{\partial ^{6}\bar{%
\partial}\bar{Q}_{5}\Theta }\right.  \\ 
\\ 
\left. +49F_{n}^{\partial Q_{5}\bar{\partial}\bar{Q}_{5}\Theta
}+10iF_{n}^{\partial ^{6}\bar{R}_{6}\Theta }+35iF_{n}^{\partial Q_{5}\bar{%
\partial}^{6}\Theta }+25iF_{n}^{R_{6}\bar{\partial}^{6}\Theta }\right) +%
\frac{1}{35m^{10}\langle \Theta \rangle }\left( -33F_{n}^{\partial ^{5}\bar{%
\partial}^{5}\Theta }+14F_{n}^{Q_{5}\bar{Q}_{5}\Theta }\right.  \\ 
\\ 
\left. +26iF_{n}^{\partial ^{5}\bar{Q}_{5}\Theta }-33iF_{n}^{Q_{5}\bar{%
\partial}^{5}\Theta }\right) +\frac{1}{5m^{8}\langle \Theta \rangle }\left(
-3F_{n}^{R_{4}\bar{R}_{4}\Theta }-2iF_{n}^{\partial ^{4}\bar{R}_{4}\Theta
}+5iF_{n}^{R_{4}\bar{\partial}^{4}\Theta }\right) +\frac{1}{m^{6}\langle
\Theta \rangle }F_{n}^{\partial ^{3}\bar{\partial}^{3}\Theta } \\ 
\\ 
+\frac{10}{m^{4}\langle \Theta \rangle }F_{n}^{\partial ^{2}\bar{\partial}%
^{2}\Theta }-\frac{20}{m^{2}\langle \Theta \rangle }F_{n}^{\partial \bar{%
\partial}\Theta }+\frac{10}{\langle \Theta \rangle }F_{n}^{\Theta }+\frac{1}{%
5m^{8}\langle \Theta \rangle ^{2}}\left( iF_{n}^{\partial ^{4}\bar{S}_{4}T%
\bar{T}}-F_{n}^{S_{4}\bar{S}_{4}T\bar{T}}\right) +\frac{5}{m^{4}\langle
\Theta \rangle ^{2}}F_{n}^{\partial ^{2}\bar{\partial}^{2}T\bar{T}} \\ 
\\ 
-\frac{10}{m^{2}\langle \Theta \rangle ^{2}}F_{n}^{\partial \bar{\partial}T%
\bar{T}}+\frac{5}{\langle \Theta \rangle ^{2}}F_{n}^{T\bar{T}}
\end{array}
\end{equation}
\vspace{0.07in}\vspace{0.07in}\vspace{0.07in}\vspace{0.07in} 
\begin{equation}
\begin{array}{l}
F_{n}^{E,K_{4}}\text{ }=\frac{1}{35m^{12}\langle \Theta \rangle }\left(
-12F_{n}^{\partial ^{6}\bar{\partial}^{6}\Theta }+7iF_{n}^{\partial ^{6}\bar{%
\partial}\bar{Q}_{5}\Theta }-5iF_{n}^{\partial ^{6}\bar{R}_{6}\Theta
}\right) +\frac{1}{35m^{10}\langle \Theta \rangle }\left( 33F_{n}^{\partial
^{5}\bar{\partial}^{5}\Theta }+7iF_{n}^{\partial ^{5}\bar{Q}_{5}\Theta
}\right)  \\ 
\\ 
+\frac{1}{5m^{8}\langle \Theta \rangle }\left( -7F_{n}^{\partial ^{4}\bar{%
\partial}^{4}\Theta }-3iF_{n}^{\partial ^{4}\bar{R}_{4}\Theta }\right) +%
\frac{1}{m^{6}\langle \Theta \rangle }F_{n}^{\partial ^{3}\bar{\partial}%
^{3}\Theta }+\frac{1}{5m^{8}\langle \Theta \rangle ^{2}}\left(
F_{n}^{\partial ^{4}\bar{\partial}^{4}T\bar{T}}-iF_{n}^{\partial ^{4}\bar{S}%
_{4}T\bar{T}}\right) 
\end{array}
\end{equation}
\vspace{0.07in}\vspace{0.07in}\vspace{0.07in}\vspace{0.07in} 
\begin{equation}
\begin{array}{l}
F_{n}^{K_{5}}\text{ }=\frac{1}{1225m^{12}\langle \Theta \rangle }\left(
-94F_{n}^{\partial ^{6}\bar{\partial}^{6}\Theta }+25F_{n}^{R_{6}\bar{R}%
_{6}\Theta }+35F_{n}^{R_{6}\bar{\partial}\bar{Q}_{5}\Theta
}+35F_{n}^{\partial Q_{5}\bar{R}_{6}\Theta }+84iF_{n}^{\partial ^{6}\bar{%
\partial}\bar{Q}_{5}\Theta }\right.  \\ 
\\ 
\left. -49F_{n}^{\partial Q_{5}\bar{\partial}\bar{Q}_{5}\Theta
}-10iF_{n}^{\partial ^{6}\bar{R}_{6}\Theta }-84iF_{n}^{\partial Q_{5}\bar{%
\partial}^{6}\Theta }+10iF_{n}^{R_{6}\bar{\partial}^{6}\Theta }\right) +%
\frac{1}{175m^{10}\langle \Theta \rangle }\left( 66F_{n}^{\partial ^{5}\bar{%
\partial}^{5}\Theta }-14F_{n}^{Q_{5}\bar{Q}_{5}\Theta }\right.  \\ 
\\ 
\left. -26iF_{n}^{\partial ^{5}\bar{Q}_{5}\Theta }+26iF_{n}^{Q_{5}\bar{%
\partial}^{5}\Theta }\right) +\frac{1}{25m^{8}\langle \Theta \rangle }\left(
-7F_{n}^{\partial ^{4}\bar{\partial}^{4}\Theta }+3F_{n}^{R_{4}\bar{R}%
_{4}\Theta }+2iF_{n}^{\partial ^{4}\bar{R}_{4}\Theta }-2iF_{n}^{R_{4}\bar{%
\partial}^{4}\Theta }\right)  \\ 
\\ 
-\frac{2}{m^{4}\langle \Theta \rangle }F_{n}^{\partial ^{2}\bar{\partial}%
^{2}\Theta }+\frac{4}{m^{2}\langle \Theta \rangle }F_{n}^{\partial \bar{%
\partial}\Theta }-\frac{2}{\langle \Theta \rangle }F_{n}^{\Theta }+\frac{1}{%
25m^{8}\langle \Theta \rangle ^{2}}\left( F_{n}^{\partial ^{4}\bar{\partial}%
^{4}T\bar{T}}+F_{n}^{S_{4}\bar{S}_{4}T\bar{T}}+iF_{n}^{S_{4}\bar{\partial}%
^{4}T\bar{T}}\right.  \\ 
\\ 
\left. -iF_{n}^{\partial ^{4}\bar{S}_{4}T\bar{T}}\right) -\frac{1}{%
m^{4}\langle \Theta \rangle ^{2}}F_{n}^{\partial ^{2}\bar{\partial}^{2}T\bar{%
T}}+\frac{2}{m^{2}\langle \Theta \rangle ^{2}}F_{n}^{\partial \bar{\partial}T%
\bar{T}}-\frac{1}{\langle \Theta \rangle ^{2}}F_{n}^{T\bar{T}}
\end{array}
\end{equation}
\begin{equation*}
\end{equation*}
for the kernels arising at level $(6,6)$.

\end{document}